\documentclass{aastex}          
\usepackage{spr-astr-addons}    
\usepackage{url}
\usepackage{graphicx}

\usepackage{times,natbib}
\usepackage{color,soul}
\usepackage[colorlinks = true,linkcolor = blue,urlcolor  = blue,citecolor = blue,anchorcolor = blue]{hyperref}

\DeclareRobustCommand{\ion}[2]{%
\relax\ifmmode
\ifx\testbx\f@series
{\mathbf{#1\,\mathsc{#2}}}\else
{\mathrm{#1\,\mathsc{#2}}}\fi
\else\textup{#1\,{\mdseries\textsc{#2}}}%
\fi}

\usepackage{float}

\newcommand{\lzifu} {{\scshape lzifu}}
\newcommand{\ppxf} {{\scshape ppxf}}
\newcommand{\idl} {{\scshape idl}}

\begin{document}
%
\title{LZIFU: an emission-line fitting toolkit for integral field spectroscopy data}

\shorttitle{LZIFU}
\shortauthors{Ho et al. }

\author{I-Ting Ho\altaffilmark{1,2}} 
\affil{itho@ifa.hawaii.edu}
\and 
\author{Anne M. Medling\altaffilmark{2}}
\and 
\author{Brent Groves\altaffilmark{2}}
\and
\author{Jeffrey A. Rich\altaffilmark{3,4}}
\and 
\author{David S. N. Rupke\altaffilmark{5}}
\and
\author{Elise Hampton\altaffilmark{2}}
\and
\author{Lisa J. Kewley\altaffilmark{1,2}}
\and
\author{Joss Bland-Hawthorn\altaffilmark{6}}
\and
\author{Scott M. Croom\altaffilmark{6,7}}
\and
\author{Samuel Richards\altaffilmark{6,7,8}}
\and
\author{Adam L. Schaefer\altaffilmark{6,7,8}}
\and
\author{Rob Sharp\altaffilmark{2,7}}
\and
\author{Sarah M. Sweet\altaffilmark{2}}

\email{itho@ifa.hawaii.edu}
\altaffiltext{1}{Institute for Astronomy, University of Hawaii, 2680 Woodlawn Dr., Honolulu, HI 96822, USA}
\altaffiltext{2}{Research School of Astronomy and Astrophysics, Australian National University, Cotter Rd., Weston ACT 2611, Australia}
\altaffiltext{3}{Infrared Processing and Analysis Center, California Institute of Technology, 1200 E. California Blvd., Pasadena, CA 91125, USA}
\altaffiltext{4}{Observatories of the Carnegie Institution of Washington, 813 Santa Barbara St., Pasadena, CA 91101, USA}
\altaffiltext{5}{Department of Physics, Rhodes College, Memphis, TN 38112, USA}
\altaffiltext{6}{Sydney Institute for Astronomy, School of Physics, University of Sydney, NSW 2006, Australia}
\altaffiltext{7}{ARC Centre of Excellence for All-sky Astrophysics (CAASTRO)}
\altaffiltext{8}{Australian Astronomical Observatory, PO Box 915, North Ryde NSW 1670, Australia}

\begin{abstract}
We present \lzifu\ (LaZy-IFU), an \idl\ toolkit for fitting multiple emission lines simultaneously in integral field spectroscopy (IFS) data. \lzifu\ is useful for the investigation of the dynamical, physical and chemical properties of gas in galaxies. \lzifu\ has already been applied to many world-class IFS instruments and large IFS surveys, including the Wide Field Spectrograph, the new Multi Unit Spectroscopic Explorer (MUSE), the Calar Alto Legacy Integral Field Area (CALIFA) survey, the Sydney-Australian-astronomical-observatory Multi-object Integral-field spectrograph (SAMI) Galaxy Survey. Here we describe in detail the structure of the toolkit, and how the line fluxes and flux uncertainties are determined, including the possibility of having multiple distinct kinematic components. We quantify the performance of \lzifu, demonstrating its accuracy and robustness.We also show examples of applying \lzifu\ to CALIFA and SAMI data to construct emission line and kinematic maps, and investigate complex, skewed line profiles presented in IFS data. The code is made available to the astronomy community through github. \lzifu\ will be further developed over time to other IFS instruments, and to provide even more accurate line and uncertainty estimates. 
\end{abstract}

\section{Introduction}\label{ho16b-sec1}
Galaxy emission-line spectroscopy has always been a powerful tool for the analysis of the dynamical, physical and chemical properties of galaxies. Traditionally, spectroscopy of galaxies has been obtained by dispersing the light either across a slit (sacrificing one spatial dimension) or from a fibre (producing a single integrated spectrum). Active development of modern integral field spectroscopy (IFS) has made capturing 3-dimensional structures of galaxies very efficient, revolutionising the way we observe and study galaxies.

The complex and perhaps stochastic nature of different physical processes governing galaxy evolution has inspired large galaxy surveys. In recent decades, large fibre and slit spectroscopic surveys such as the Sloan Digital Sky Survey \citep[SDSS;][]{York:2000qe}, the 2dF Galaxy Redshift Survey \citep[][]{Colless:1999kl}, and the Deep Extragalactic Evolutionary Probe 2 survey \citep[DEEP2;][]{Davis:2003lr} have drastically improved our understanding of the global (unresolved) properties of galaxy populations at different epochs of the Universe. Integral field spectroscopic surveys have recently become feasible, providing access simultaneously to both spectral and kinematic information of large numbers of galaxies. Two pioneering IFS surveys, the SAURON survey \citep{Bacon:2001rt} and its extension the $\mathrm{ATLAS^{3D}}$ survey \citep{Cappellari:2011ys}, studied about 260 early type galaxies in the local Universe ($z<0.01$). Surveys targeting both the blue and red galaxy populations, such as the Calar Alto Legacy Integral Field Area (CALIFA) survey \citep{Sanchez:2012fj}, the Sydney-Australian-astronomical-observatory Multi-object Integral-field spectrograph (SAMI) Galaxy Survey \citep{Croom:2012qy,Bryant:2015bh}, and the Mapping Nearby Galaxies at Apache Point Observatory survey \citep[MaNGA;][]{Bundy:2015kx}, are currently underway. These IFS surveys will provide critical information to bridge the knowledge gaps resulting from the limited spatial and kinematic information delivered by previous single-fibre and slit spectroscopic surveys. 

The sample sizes and data flows of these modern IFS surveys are substantial. With each data cube containing typically one to two thousand spectra, the CALIFA survey plans to observe about 600 galaxies in the local Universe ($0.005<z<0.03$); the SAMI Galaxy Survey will reach a sample size of 3,400 galaxies at $z<0.12$; and the MaNGA survey will build up a sample of 10,000 galaxies at a similar redshift to the SAMI Galaxy Survey. Future surveys using high-multiplex integral field unit (IFU) instrument such as HECTOR on the Anglo-Australian Telescope will observe on the order of 100,000 galaxies \citep{Lawrence:2012fk,Bland-Hawthorn:2015dn}. Current and forthcoming wide-field IFU instruments are also delivering large quantity of high quality data, such as the Wide Field Spectrograph (WiFeS) on the Australian National University 2.3-m telescope \citep{Dopita:2007kx,Dopita:2010yq}, the new Multi Unit Spectroscopic Explorer (MUSE) on the Very Large Telescope \citep{Bacon:2010ph}, the SITELLE instrument on the Canada France Hawaii Telescope \citep{Grandmont:2012sp}, the Keck Cosmic Web Imager at the W. M. Keck Observatory \citep{Martin:2010th,Morrissey:2012ij}.

Significant efforts have been placed in developing corresponding tools for analysing large volume of spectroscopic data. The stellar continuum contains valuable information about the stellar kinematics, chemistry and star formation history of galaxies. Packages such as the STEllar Content via Maximum A Posteriori \citep[{\scshape stecmap};][]{Ocvirk:2006fj} package, the penalized pixel-fitting \citep[\ppxf;][]{Cappellari:2004uq} routine and the {\scshape starlight} package \citep{Cid-Fernandes:2005fk} can perform spectral template fitting and extract various stellar properties.  For investigating gas physics, the emission lines fitting tools such as the Gas AND Absorption Line Fitting code \citep[{\scshape gandalf};][]{Sarzi:2006lr}, the {\scshape fit3d} package (\citealt{Sanchez:2006uq,Sanchez:2007qy}; and the successor {\scshape pipe3d}; \citealt{Sanchez:2016fk,Sanchez:2016lr}), and the Peak ANalysis utility \citep[{\scshape pan}\footnote{{\scshape pan} was subsequently adapted and modified by Mark Westmoquette for astronomical requirements. See \url{http://ifs.wikidot.com/pan}. };][]{dimeo2005}  are commonly adopted to measure emission line fluxes and kinematics. 

As the spectral resolution of the instruments continue to improve, the intrinsic non-Gaussian line profile complicates the emission line analysis. When the spectral resolution is high ($\rm R > 3000 $), galaxies with active gas dynamics, such as winds, outflows or AGN, usually present skewed line profiles that require fitting multiple, assumed Gaussian, components to separate the different kinematic components overlapping in the line-of-sight direction (also referred as ``spectral decomposition''). Performing spectral decomposition on large datasets is non-trivial as significant human input is usually required. Here, we present our emission line fitting pipeline LaZy-IFU\footnote{The framework of the \lzifu\  stems from {\scshape uhspecfit}, a tool developed at the University of Hawai\textquoteright i and employed in several previous spectroscopic studies on gas abundances and outflows \citep[e.g.,][]{Zahid:2011fj,Rupke:2010fk,Rich:2010yq,Rich:2012oq,Rupke:2011fk}.} (\lzifu; written in the Interactive Data Language [{\idl}]), which is designed to eliminate the need for individual treatment of each of many thousands of spectra across an IFS galaxy survey (such as CALIFA, SAMI or MaNGA). 

The main objective of \lzifu\ is to extract 2-dimensional emission line flux maps and kinematic maps useful for investigating gas physics in galaxies. \lzifu\ has already been adopted in various publications using data from multiple instruments and surveys, including MUSE \citep[][]{Kreckel:2016zl}, SAMI \citep[e.g.][]{Ho:2014uq,Richards:2014lr,Allen:2015dk,Ho:2016rf}, CALIFA \citep{Davies:2014fy,Ho:2015hl}, WiFeS \citep{Ho:2015hl,Dopita:2015lq,Dopita:2015eu,Vogt:2015ec,Medling:2015eu}, and SPIRAL on the Anglo-Australian Telescope \citep{McElroy:2015ve}. The following characteristics were considered carefully while developing \lzifu. First, the pipeline must perform spectral decomposition automatically without needing repeated human instructions. Second, the pipeline needs to be scriptable for batch reduction, such that when necessary the same results can be reproduced by re-executing the same scripts. Third, the pipeline must be flexible and generalised so that data from most modern IFS instruments can be accepted without major restructuring of the inputs. Finally, the calculation speed must be optimised and the pipeline has to take advantage of parallel processing because of 1) the huge data flow from multiplexed IFS surveys, and 2) the possibility of fitting the same datasets multiple times for various experimental purposes.

The focus of this paper is to present the core structure of \lzifu\ (Section~\ref{ho16b-sec2}), and examine the errors produced by the pipeline (Section~\ref{ho16b-sec4}). We also show examples of applying \lzifu\ on the CALIFA survey and SAMI Galaxy survey (Section~\ref{ho16b-sec3}). Finally, the code will be continuously maintained and made available to the public through github (\url{https://github.com/hoiting/LZIFU/releases}). We discuss future plans for the code in Section~\ref{ho16b-sec5}.

\section{\lzifu: The spectral fitting toolkit}\label{ho16b-sec2}
\subsection{Overview}\label{ho16b-sec2.1}

To arrive at 2D maps of line fluxes, velocity and velocity dispersion, \lzifu\ first removes the continuum before modelling user-assigned emission line(s) on a spaxel-to-spaxel basis. If tailored continuum models already exist, users have the option of directly subtracting the continuum by feeding \lzifu\ the continuum models in Flexible Image Transport System (FITS) format. The subsequent emission line fitting follows the Levenberg-Marquardt least-square method to find the most probable models (with maximum likelihood) describing the emission line spectra. Each emission line can be modelled by up to 3 Gaussians describing (potentially) different kinematic components. The final products delivered by \lzifu\ are continuum cubes, emission line cubes, emission line flux (and corresponding error) maps, and kinematic (and corresponding error) maps stored in multi-extension FITS files. 

For historical reasons, \lzifu\ was originally designed for two-sided IFS data with each object having one blue and one red data cube. The two data cubes can have different spectral resolutions, but are required to cover {\it non-overlapping} spectral ranges. Such an instrumental setup is common in instrument designs and large area IFS surveys because one can achieve a trade-off between spectral coverage and spectral resolution, given that the numbers of CCD pixels are always limited. To generalise the application of \lzifu, the pipeline was modified later to accept one-sided IFS data by disabling procedures related to the blue data. Below, we elaborate on the continuum fitting and emission line fitting procedures based on two-sided data. 

\subsection{Continuum fitting}\label{ho16b-sec2.2}

\begin{figure*}[!t]
\centering
\includegraphics[width = \textwidth]{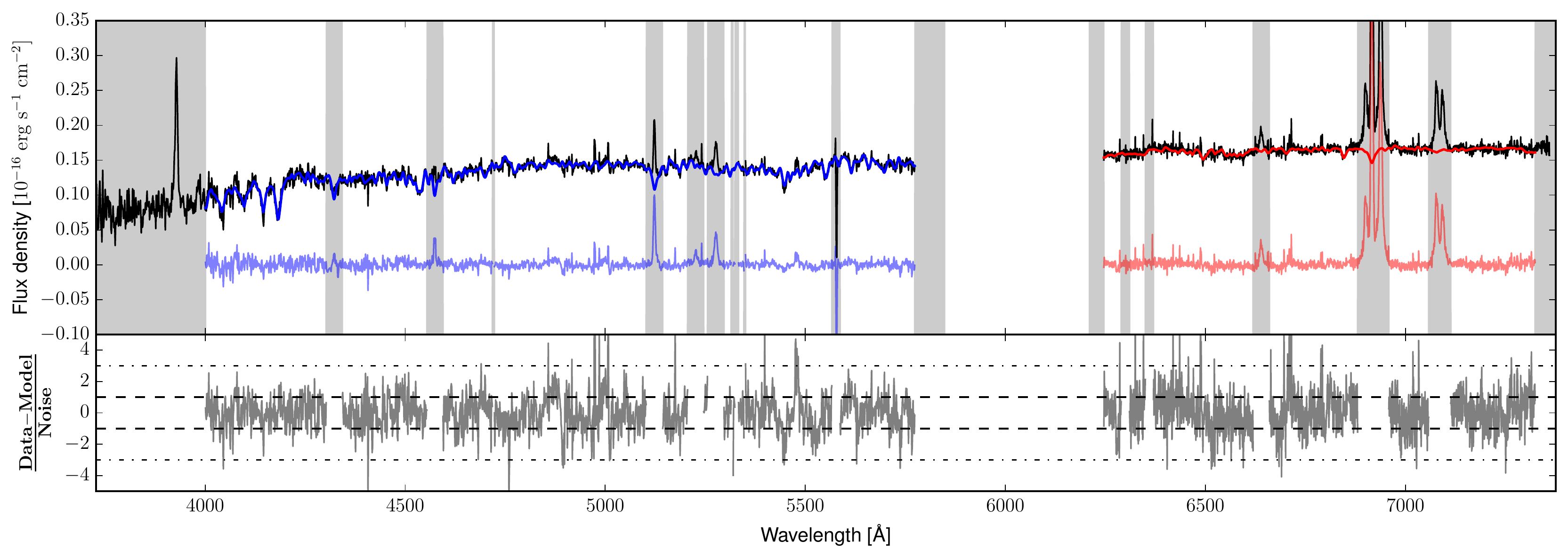}
\caption{An example of \ppxf\ continuum fitting using data from the SAMI Galaxy Survey. Channels not included in the continuum fitting are masked by grey bands (i.e., bad pixels, cosmic rays, poorly-subtracted sky emission lines, and nebular emission lines). Continuum subtracted data (lower blue and red lines) are subsequently used to perform emission line fitting. }\label{ho16b-fig1}
\end{figure*}

When pre-determined continuum models are not provided by the users, \lzifu\ models the continuum using the penalised pixel-fitting routine \citep[\ppxf;][]{Cappellari:2004uq} that performs fits the underlying absorption continuum using a series of input spectral templates from stars or modeled simple stellar populations (SSPs) convolved with a parameterized velocity distribution. The \ppxf\ routine is wrapped in \lzifu\ as the default continuum fitting method. In our implementation, the data and spectral templates are first aligned and rectified to the same spectral characteristics (i.e. wavelength coverage, spectral resolution, and channel width) before fitting the continuum. A combined spectrum (of the blue and red data) is formed for each spaxel by convolving the data to a common spectral resolution, and resampling the data onto a common spectral grid. The data cube with poorer spectral resolution determines the spectral resolution and channel width of the combined spectrum. Various SSP templates collected from the literature are included in \lzifu\ as \idl\ .sav files, so the users can directly select the preferred library of SSP models. The selected SSP templates are redshifted, spectrally trimmed, and spectrally convolved to match the combined spectrum. To fit the underlying absorption-line spectrum, \ppxf\ compares linear combinations of the SSP models with the combined spectrum in a least-square sense, during which the stellar velocity dispersion, stellar velocity, and reddening are constrained simultaneously. Channels contaminated by night sky emission lines and nebular emission lines from the galaxy are masked out prior to the fitting. Poorly-subtracted sky emission lines, with other defect channels, can be masked by providing external masks that specify the wavelength intervals to ignore. Users are also required to specify the emission lines and the width around the emission lines that should be excluded from the continuum fit. Our custom implementation of \ppxf\ allows the users to control critical \ppxf\ keywords directly from a \lzifu\ setup file. Other hardwired functionalities of \ppxf\ can be adapted for different applications by modifying the \lzifu\ source code. After the best solution of spectral fitting is found, the continuum models are reconstructed separately for both sides of the data at their native resolutions. The advantage of this implementation is twofold: we utilise the largest possible spectral coverage to constrain the spectral fitting solution, and the reconstructed continuum models retain the original spectral resolution of the data.

\bigskip

Systematic errors in SSP models, residual calibration errors, and potential power law continuum from non-stellar components often cause spectral fitting routines to fail to achieve a perfect description of a spectrum, which would be characterised by a reduced-$\chi^2$ ($\chi_{\nu}^2$) of approximately 1. To account for these systematic errors and possible non-stellar contributions, a polynomial term can be implemented. In \ppxf, additive or multiplicative Legendre polynomials can be included and fit simultaneously with the spectral templates. These options are also maintained and passed on to \ppxf. In some situations, the users may wish not to fit polynomials simultaneously with the spectral templates to avoid the continuum fit becoming highly degenerate. In these cases, the continuum subtracted-spectra may not be flat, which can affect the subsequent line flux measurements. To further flatten the continuum-subtracted spectra, \lzifu\ provides an extra option of fitting Legendre polynomials separately to the continuum-subtracted blue and red spectra. Here, the least-square fitting is performed using the {\scshape bvls} (bounded-value least-square) algorithm developed by \citet{Lawson:1974vn} and implemented in \idl\ by Michele Cappellari.

The principal objective of the custom implementation of \ppxf\ is to correct for stellar absorption features affecting predominately the Balmer lines, and to remove the stellar continuum, such that gas physics can be derived from fitting emission lines to continuum-free spectra. The goal is not to constrain stellar parameters such as the stellar population, age and metallicity, which are known to be highly degenerate and require careful investigation of numerous local minima in the $\chi^2$ space \citep[e.g.,][]{Cid-Fernandes:2014qy}.

In Figure~\ref{ho16b-fig1}, we show an example of the continuum fit of one spectrum from the SAMI Galaxy Survey. The blue and red data have  $1\sigma$ spectral resolutions of 1.15 and 0.72~\AA, respectively; and we use the SSP spectral libraries constructed by \citet{Gonzalez-Delgado:2005lr} with an additional Legendre polynomial of up to 12 order of Legendre polynomials to fit the continuum. Channels affected by bad pixels, cosmic rays, strong sky lines, or nebular emission lines are masked by grey bands, and are not considered in the continuum fit.

\subsection{Emission line fitting}\label{ho16b-sec2.3}

\begin{figure}[!hb]
\centering
\plotone{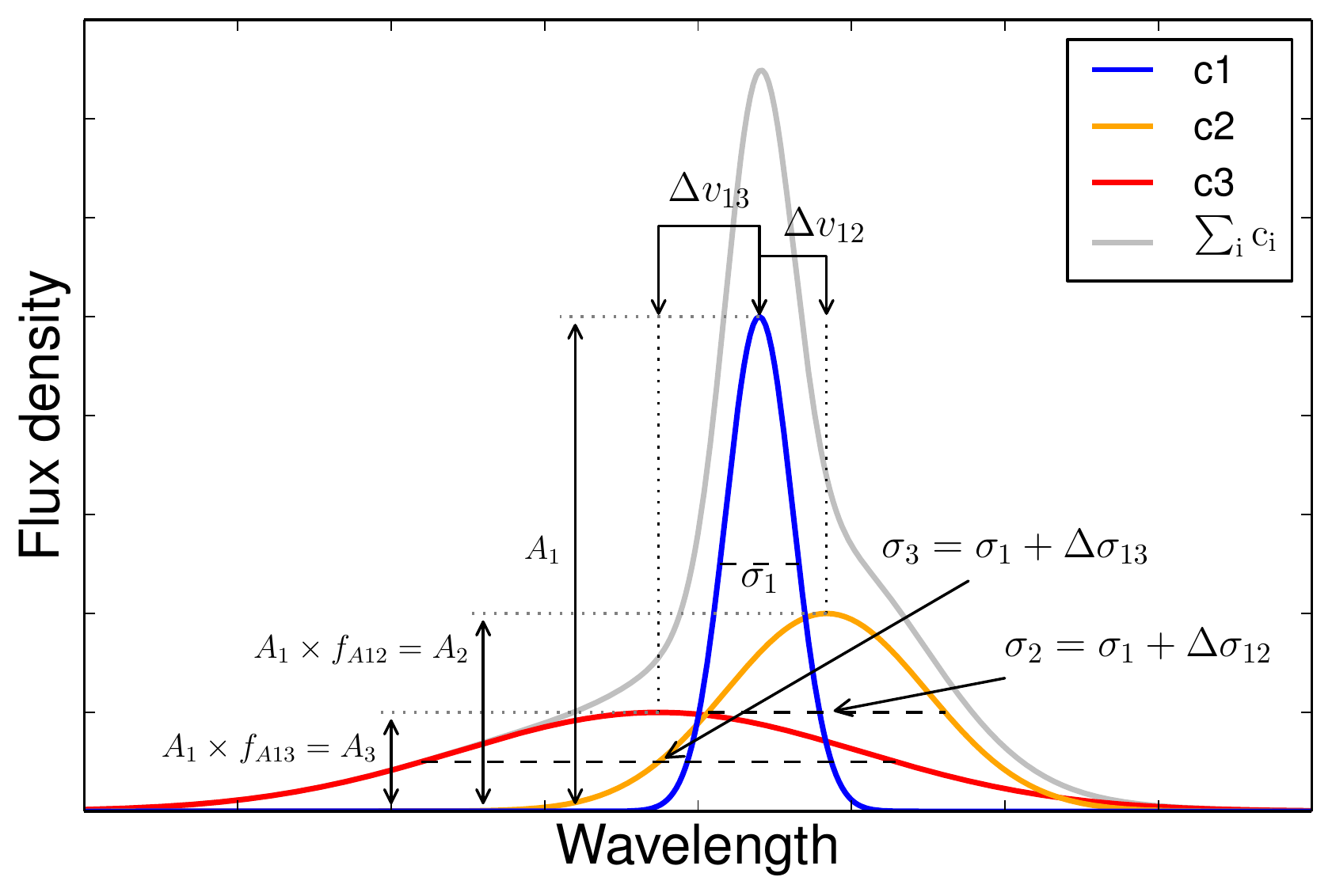}
\caption{A schematic illustrating the definition of the two sets of parameters $(f_{A12}, \Delta v_{12}, \Delta\sigma_{12})$ and $(f_{A13}, \Delta v_{13}, \Delta\sigma_{13})$ controlling the initial guesses of the second {\it c2} (intermediate) and the third {\it c3} (broad) kinematic component relative to the first (narrow) kinematic component {\it c1}. }\label{ho16b-fig2}
\end{figure}

\begin{figure*}
\centering
\includegraphics[width = \textwidth]{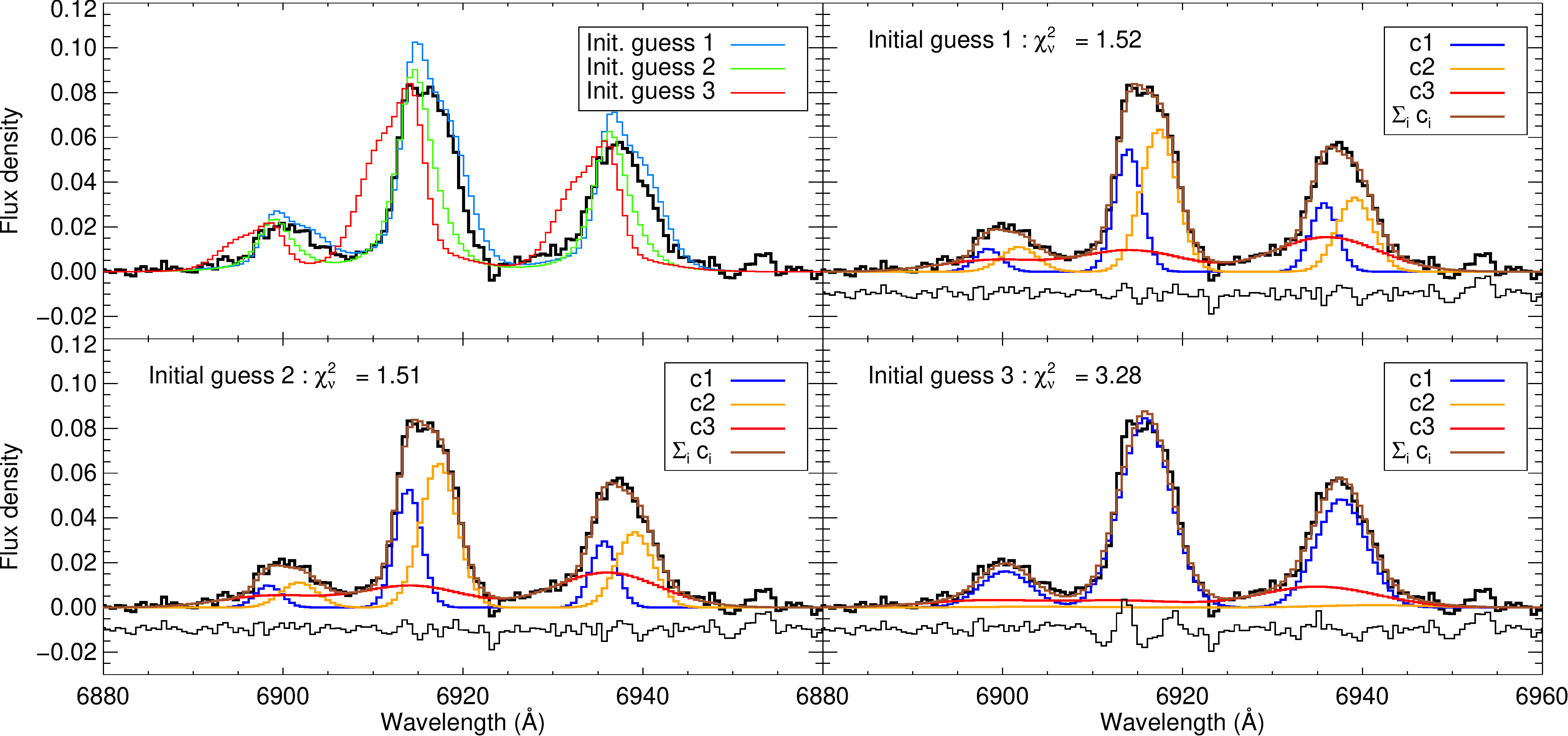}
\caption{Demonstration of the importance of using multiple initial guesses to reject local minima. The data (thick black lines; from \citealt{{Ho:2014uq}}) are fit with three different initial guesses to perform 3-component fitting to the {[\ion{N}{ii}]~$\lambda\lambda$6548,83} and H$\alpha$ lines. The three different initial guesses are shown in the upper right panel (see Section~\ref{ho16b-sec2.3.1}). The best-fits are shown in the other three panels, with the resulting reduced-$\chi^2$ ($\chi_{\nu}^2$) labeled in each panel. The first and the second initial guesses yield very similar fits and $\chi_{\nu}^2$, but the third initial guess arrives a very different solution with much worse $\chi_{\nu}^2$ (clearly visible from the residuals shown as thin black lines). 
}\label{ho16b-fig3}
\end{figure*}

Emission lines are fit in the continuum subtracted spectra. The lines are assumed to be gaussian in shape and are fit as Gaussians using the Levenberg-Marquardt least-square method implemented in {\scshape idl} \citep[{\scshape mpfit};][]{Markwardt:2009lr}. Users have the option of fitting the emission lines using multiple Gaussian components, with this currently limited to a maximum of 3 components. All the lines are fit simultaneously with each kinematic component constrained to share the same velocity and velocity dispersion. When more than 1 component is fit, \lzifu\ sorts and groups the fitting results based on either velocity dispersions, velocities or line fluxes, and produces 2D maps of fluxes, velocity, and velocity dispersions separately for different components. Which reference value is adopted to group and sort the different components is determined by the users, and we encourage the users to consider carefully what sorting methods are best for their specific science goals. In the rest of the paper, we sort and group the components based on velocity dispersion. That is: the first component ({\it c1}) is the Gaussian fit with the narrowest velocity dispersion, and the second ({\it c2}) and third ({\it c3}) components have increasing velocity dispersions.

\subsubsection{Establishing initial guesses}\label{ho16b-sec2.3.1}

Establishing proper initial guesses for the model parameters is critical when using the Levenberg-Marquardt least-square algorithm, because the initial guesses serve as starting points for the algorithm to explore the {\it n}-dimensional $\chi^2$ space along its negative gradients. 

In \lzifu, initial guesses are established automatically by means of an internal algorithm (for the first component {\it c1}) and several external parameters determined by the user (for the second {\it c2} and third {\it c3} components). The internal algorithm searches for peak S/N in the spectrum to determine the central wavelengths and amplitudes of the first kinematic components. The redshift of the galaxy input by the user allows \lzifu\ to estimate the rough locations of the emission lines. The locations are updated if sensible stellar velocity and velocity dispersion can be obtained from the stellar continuum fit. Channels around $\rm\pm300~km~s^{-1}$ from the fiducial location of each emission line are inspected, and the channel with the highest signal-to-noise ratio (S/N) determines the amplitude guess of the first Gaussian component. When fitting multiple emission lines, the line with the best S/N anchors the initial wavelength guess of the first Gaussian component. The width of the first Gaussian component is provided by the user. For the SAMI Galaxy Survey data we typically adopt a width of $\rm 50~km~s^{-1}$. 

When fitting more than one component, how and where to place the second (and third) kinematic components are determined by a set of parameters specified by the user. We use three parameters $(f_A, \Delta v, \Delta\sigma)$ to describe the relationship between the second (or third) Gaussian component(s) and the first component. Figure~\ref{ho16b-fig2} illustrates the definitions of the parameters. The two sets of parameters $(f_{A12}, \Delta v_{12}, \Delta\sigma_{12})$ and $(f_{A13}, \Delta v_{13}, \Delta\sigma_{13})$ control the initial guesses of the second and third component, respectively. The combined profile (grey curve in Figure~\ref{ho16b-fig2}) is normalised to the peak value of the data before proceeding to solve the least-square problem.

Fitting multiple components can sometimes be sensitive to the choice of initial guesses, particularly when the S/N is poor or the spectrum is only marginally resolved. As a result, \lzifu\ allows multiple initial guesses to be generated by providing arrays of possible $f_A$, $\Delta v$ and $\Delta \sigma$ values. All possible combinations of initial guesses are solved for least-square solutions with {\scshape mpfit}, and the fit with the best minimum $\chi_{\nu}^2$ is kept as the final solution.

We demonstrate the importance of using multiple initial guesses to reject local minima in Figure~\ref{ho16b-fig3}.  Different initial guesses are adopted to model the {[\ion{N}{ii}]~$\lambda\lambda$6548,83} and H$\alpha$ lines. The spectrum comes from a galaxy observed by the SAMI Galaxy Survey and presented in \citet{Ho:2014uq}. In the upper-left panel, three different initial guesses  (color-coded lines) are generated. The second components are characterised by $f_{A12} = 1$, $\Delta\sigma_{12} = 60\rm~km~s^{-1}$, and $\Delta v_{12}$ with three possible values of $150$, $30$, and $ -150\rm~km~s^{-1}$. The initial guesses of the third component are all the same of $f_{A13} = 0.2$, $\Delta\sigma_{13} = 200\rm~km~s^{-1}$, and $\Delta v_{13} =\rm  -10~km~s^{-1}$. The first component has an initial velocity dispersion of $50\rm~km~s^{-1}$. After applying the Levenberg-Marquardt least-square algorithm, the first and second initial guesses arrive at solutions virtually indistinguishable with very similar $\chi_{\nu}^2$ of 1.51 and 1.52, respectively. The third initial guess, however, arrives at a very different solution with much higher $\chi_{\nu}^2$ of 3.28. A careful visual inspection of neighbouring spaxels reveals that there are indeed three separate kinematic components in this galaxy, but as the first and second kinematic components are spectrally close to each other in this spaxel, fitting the two narrow peaks with a single Gaussian yields a local minimum (the bottom right panel). This local minimum can be rejected by fitting with different initial guesses.

Our implementation of multiple initial guesses has several advantages. Empirical understanding of the physical characteristics of the second and third components can be incorporated directly into guiding the fits by providing proper sets of parameters ($f_A$, $\Delta v$ and $\Delta \sigma$). In principle, the {\it n-}dimensional $\chi^2$ space will be explored thoroughly if a chain of initial guesses is carefully chosen. Our algorithm trades computational expense against sensitivity to local minima. This makes the analysis of extended data sets tractable.

\subsubsection{Optional refit with smoothed initial guesses}\label{ho16b-sec2.3.2}

Optional refits are possible after fitting the data with the default initial guesses. In the refitting process, results from the 
first-pass fit to the full data cube are spatially median-smoothed to produce initial guesses to refit the data. The refitting process can be repeated multiple times. 

The reasoning behind the refitting process is that flux, velocity and velocity dispersion usually vary smoothly across the spatial dimensions, a direct result of the intrinsic properties of galaxies and the finite spatial resolution of the data. Therefore, the best fits of neighbouring spaxels contain information useful for establishing a proper initial guess. Our experience shows that the refitting process is useful for rejecting some bad results caused by ill-chosen initial guesses in the previous fits, and those poor fits triggered by the presence of local defects in the spectra (poor sky-line subtraction, uncleaned cosmic ray residuals, etc.) which force the initial guess solution into a local minima of limited relevance.

\subsection{Output}
The final products delivered by \lzifu\ are stored in multi-extension FITS files. For a more detailed description of the output data structure, the readers are directed to the readme file included in the code release package. In brief, \lzifu\ generates 3-dimensional model cubes and 2-dimensional maps from the fitting. The model cubes include both the continuum models and emission line models. All the model cubes have the same spatial and spectral dimensions as the input data cubes. These model cubes are not only useful for visualising the fits, but also for removing emission lines or continuum from the data cubes (i.e. for generating line-free or continuum-free data cubes). Emission line fluxes, velocities, and velocity dispersions (and corresponding errors) are stored in 2-dimensional maps. These 2-dimensional maps have the same spatial dimensions as the input data cubes. These maps are most useful for subsequent scientific analysis, e.g. converting line fluxes to star formation rates or extinction, producing emission line ratio maps, fitting disk models to the velocity field, etc.

\section{Applications on Survey Data}\label{ho16b-sec3}

We present three examples of flux and kinematic maps generated by \lzifu\ using public data from the CALIFA survey and the SAMI Galaxy Survey. The examples are chosen to demonstrate the use of \lzifu\ in different types of data and galaxies. The simple single component analysis is useful for dynamically stable systems or when the spectral resolution is insufficient to resolve the kinematic structures. The more complicated double and triple component analyses are required when the line profiles are skewed due to either complex gas kinematics or beam smearing. 

Although the examples below make use of CALIFA and SAMI data, \lzifu\ is not limited to these two surveys and can be adopted for any IFS data with similar characteristics, i.e. spectral coverage and resolution. Indeed, \lzifu\ has already been applied to data from multiple IFU instruments for various science cases related to gas physics. Data from the WiFeS instrument on the Australian National University 2.3-m telescope have been tested extensively with \lzifu\ \citep{Ho:2015hl,Dopita:2015lq,Dopita:2015eu,Vogt:2015ec,Medling:2015eu}. Recently, \lzifu\ has also been adopted to analyse data from MUSE (\citealt{Kreckel:2016zl}; Juneau et al. in preparation) and SPIRAL on the Anglo-Australian Telescope \citep{McElroy:2015ve}. The reader is referred to corresponding publications for more examples.

\begin{figure*}[!ht]
\centering
\includegraphics[width = \textwidth]{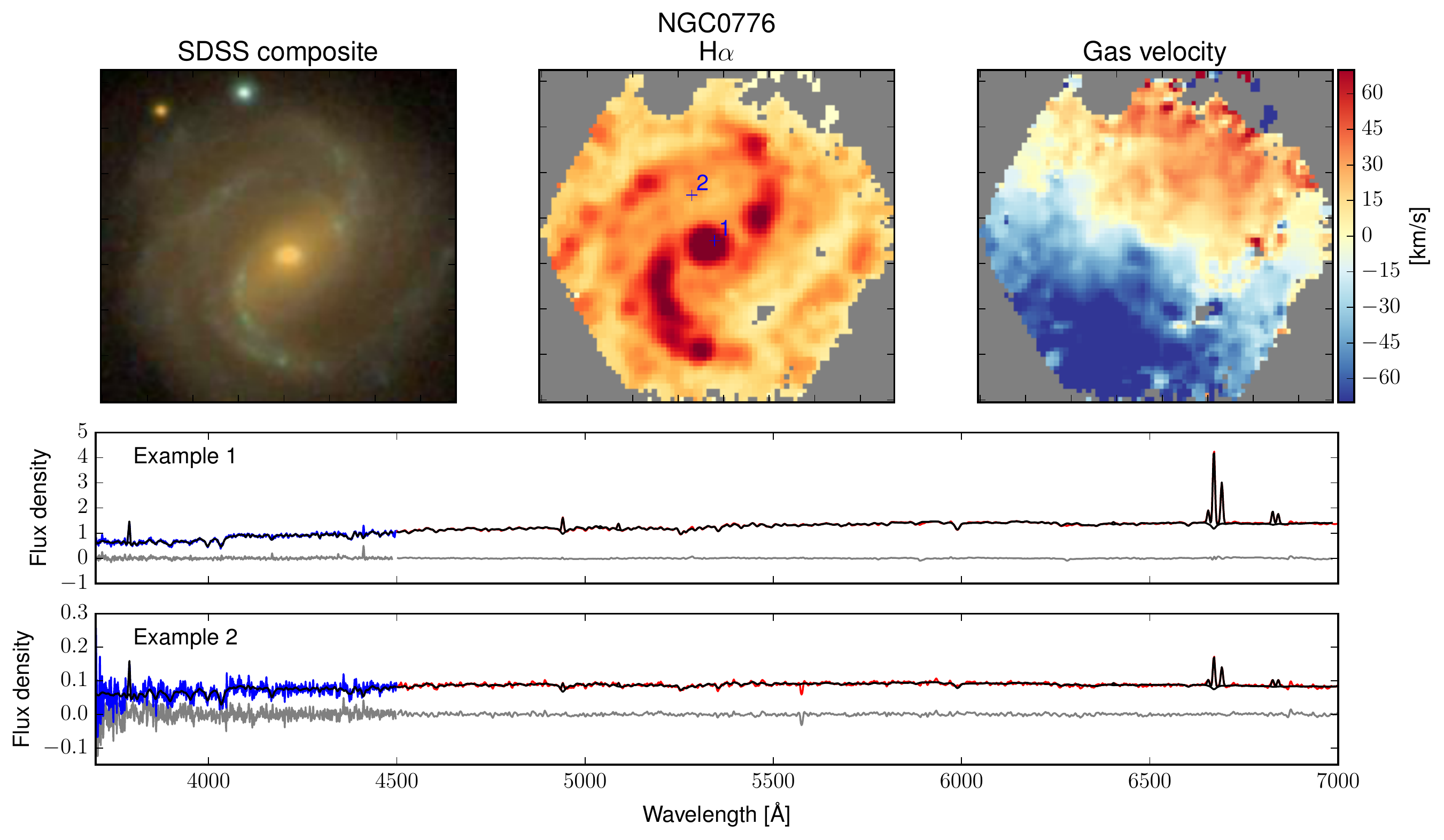}
\caption{Demonstration of applying a 1-component fit to NGC0776 using the first data release of the CALIFA survey. The SDSS {\it g,r,i} composite image, \lzifu\ H$\alpha$ and gas velocity images are shown left to right in the top row. The spectra and corresponding best-fit (continuum + line) models of the two example spaxels marked in the H$\alpha$ map are shown in the bottom two panels. We use the V1200 data (blue lines) at wavelengths smaller than 4500\AA\ and the V500 data (red lines) at wavelengths larger than 4500\AA. }\label{ho16b-fig4}
\end{figure*}

\subsection{Simple single component analysis}\label{ho16b-sec3.1}

We demonstrate 1-component fitting using data from the CALIFA survey \citep[i.e., the first data release; ][]{Husemann:2013yq}. Figure~\ref{ho16b-fig4} shows the NGC0776 SDSS  {\it g,r,i} colour composite image, H$\alpha$ map, gas velocity map, and two example spectra from the integral field data. Here, we adopt the MIUSCAT SSP libraries \citep{Vazdekis:2012yq} of solar metallicity to model the continuum. After subtracting the continuum, the line profiles appear to be simple across the entire galaxy and therefore only single component Gaussians are required to model the emission lines. The emission line maps and kinematic maps delivered by \lzifu\ are directly ready for various studies such as gas dynamics and chemical abundance.

\begin{figure}[!ht]
\centering
\plotone{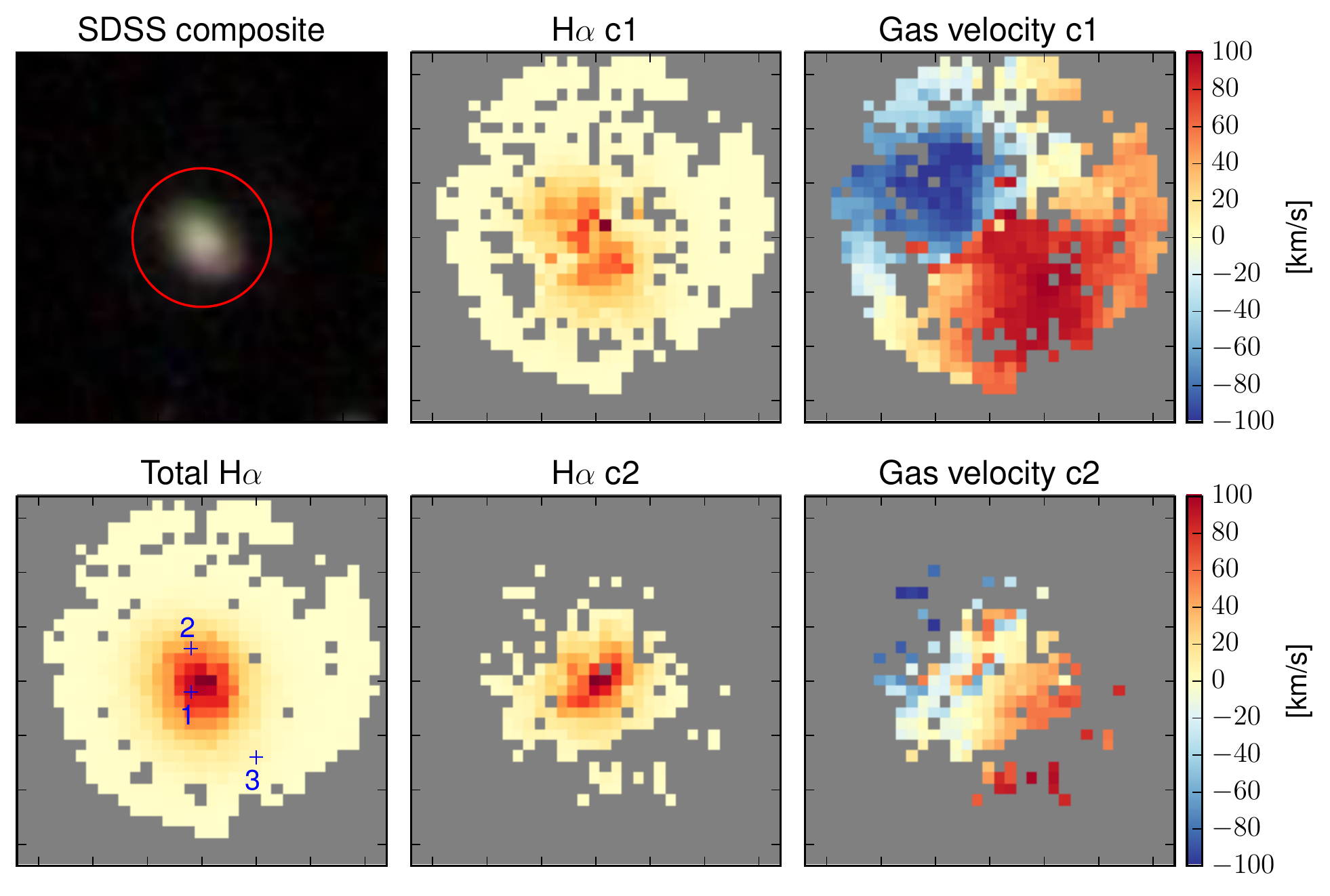}
\caption{Demonstration of multiple component fitting for a galaxy from the early data release of the SAMI Galaxy Survey (\citealt{Allen:2015lq}; GAMA ID: 594906). The SDSS {\it g,r,i} image in the upper left panel shows the circular field of view of the SAMI instrument (red circle; 15\arcsec\ in diameter). The H$\alpha$ flux and gas velocity maps of the first (narrowest component) {\it c1} and the second component {\it c2} are shown in the middle and right panels. The sum of H$\alpha$ for all components is shown in the bottom left panel (total H$\alpha$). Very few spaxels in this galaxy require the third component so the corresponding maps are not shown. The fits of the three spaxels marked in the total H$\alpha$ map are shown in Figure~\ref{ho16b-fig6}. }\label{ho16b-fig5}
\end{figure}

\begin{figure}[!ht]
\centering
\plotone{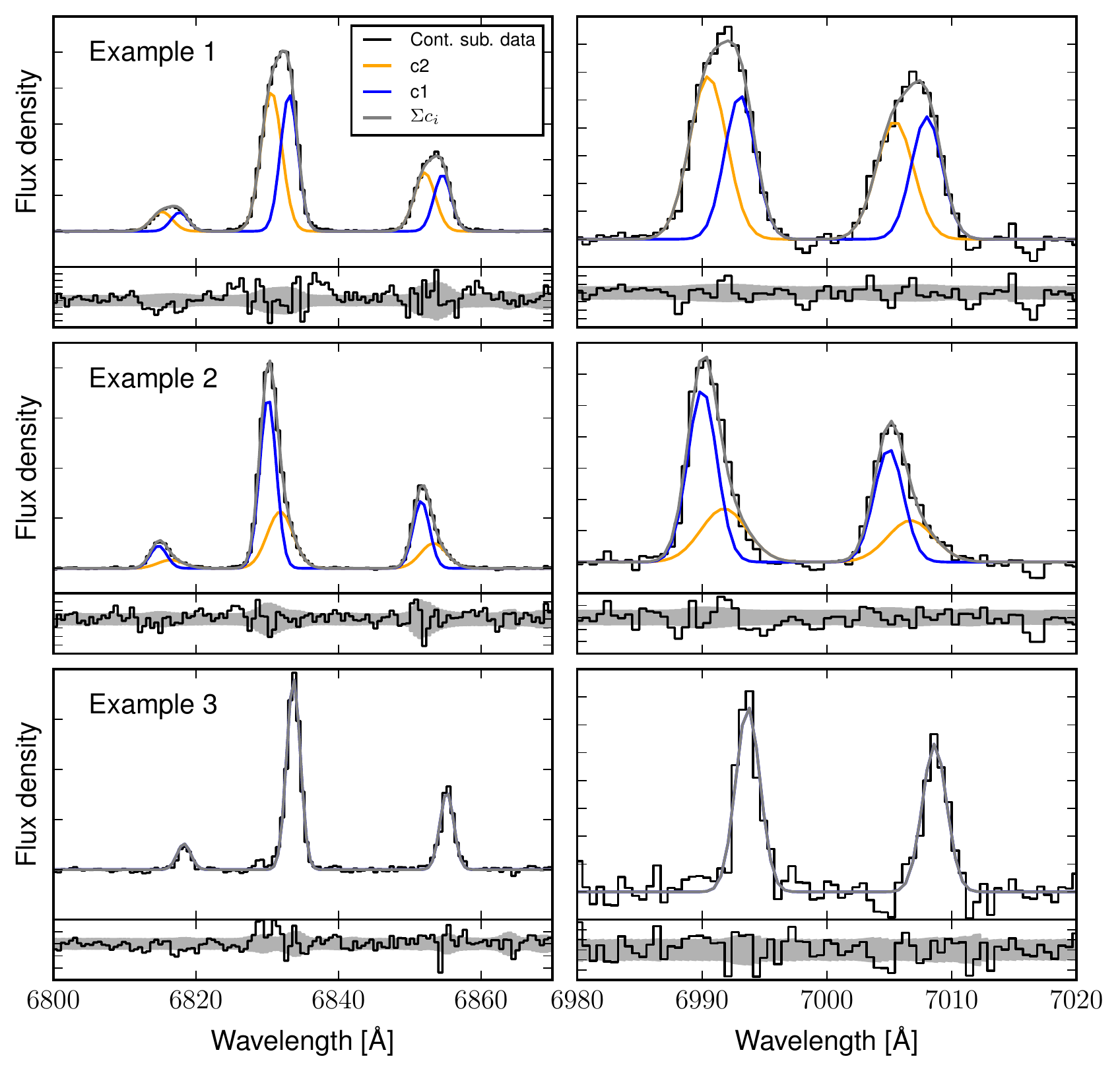}
\caption{Examples of the {[\ion{N}{ii}]~$\lambda\lambda$6548,83} + H$\alpha$ (left column) and {[\ion{S}{ii}]~$\lambda\lambda$6716,31} (right column) fits of the three spaxels marked in Figure~\ref{ho16b-fig5} (in lower left `total H$\alpha$' panel). The continuum-subtracted data and best-fit models are shown as solid lines. In the lower plot of each panel, we show the residuals as black lines, and the $\pm1\sigma$ measurement errors as grey shading. }\label{ho16b-fig6}
\end{figure}

\begin{figure}[!ht]
\centering
\plotone{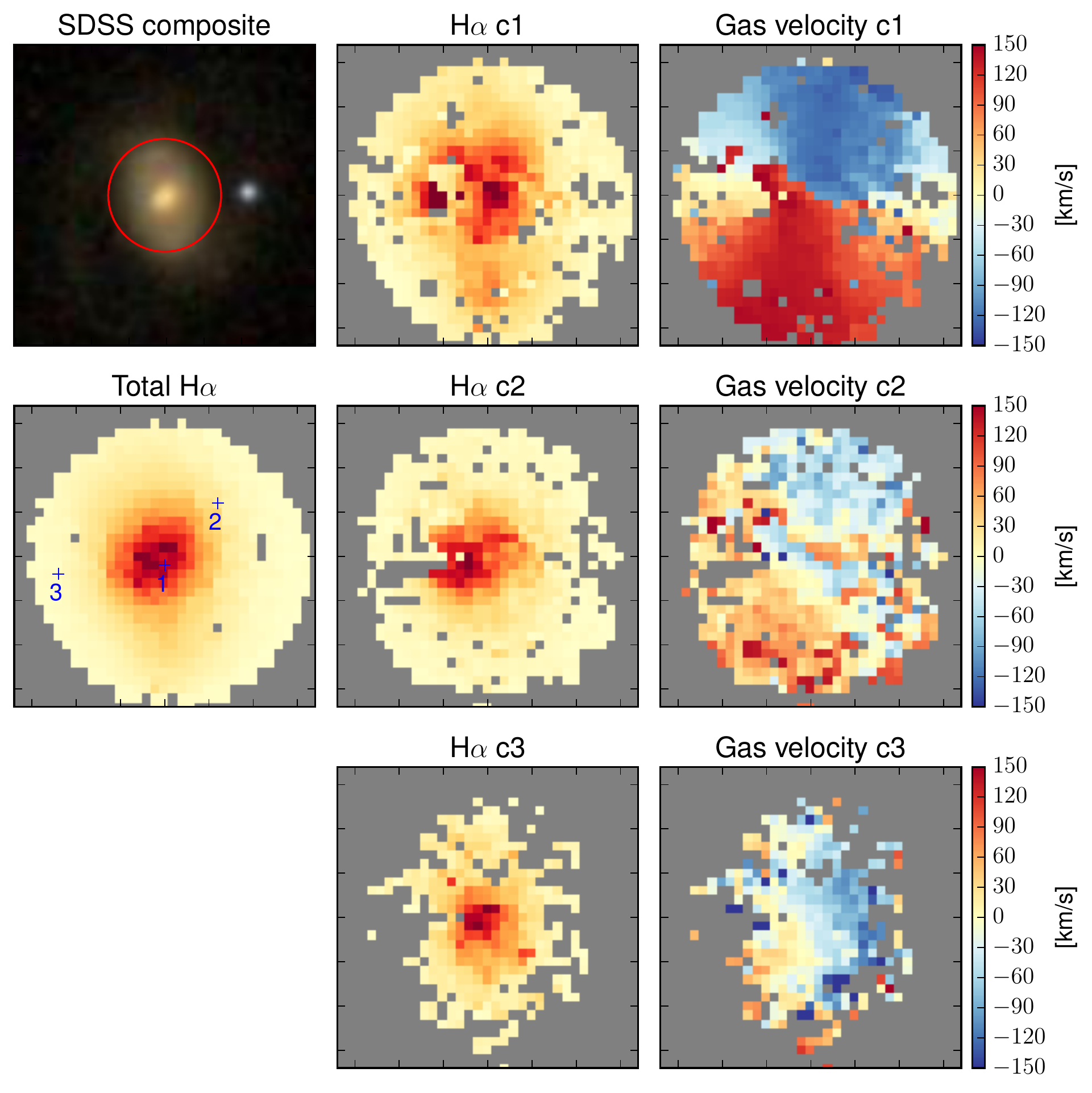}
\caption{Demonstration of multiple component fitting for a normal star-forming galaxy presented in \citet[GAMA ID: 209807]{Ho:2014uq}. The SDSS {\it g,r,i} image in the upper left panel shows the circular field of view of the SAMI instrument (red circle; 15\arcsec\ in diameter). The H$\alpha$ flux and gas velocity maps of the first (narrowest component) {\it c1}, the second component {\it c2}, and the third component {\it c3} are shown in the middle and right panels. The sum of H$\alpha$ for all components is shown in the middle left panel (total H$\alpha$). The fits of the three spaxels marked in the total H$\alpha$ map are shown in Figure~\ref{ho16b-fig6}.}\label{ho16b-fig7}
\end{figure}

\begin{figure}[!ht]
\centering
\plotone{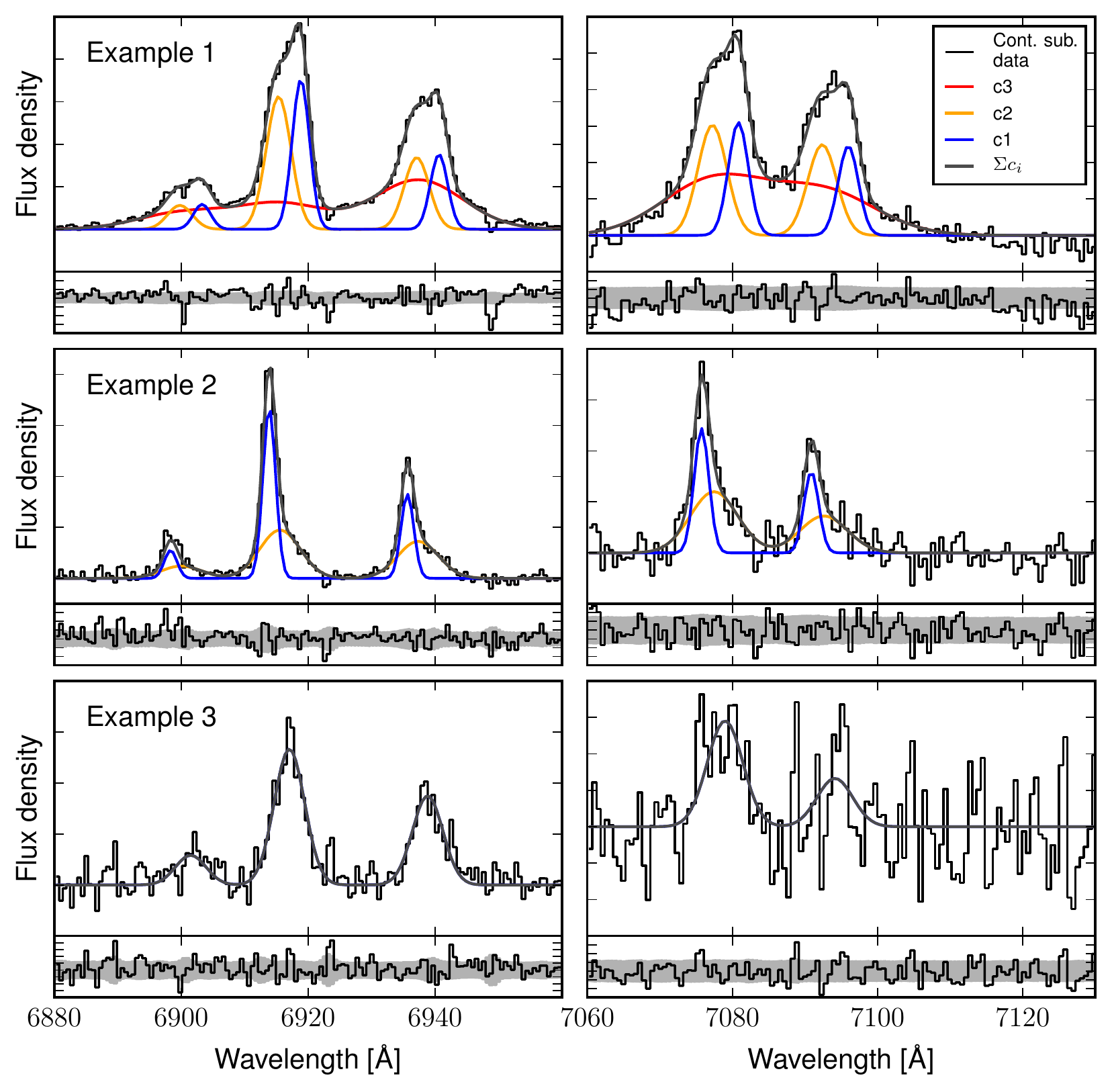}
\caption{Examples of the {[\ion{N}{ii}]~$\lambda\lambda$6548,83} + H$\alpha$ (left column) and {[\ion{S}{ii}]~$\lambda\lambda$6716,31} (right column) fits of the three spaxels marked in Figure~\ref{ho16b-fig7} (in middle left `total H$\alpha$' panel). The continuum-subtracted data and best-fit models are shown as solid lines. In the lower plot of each panel, we show the residuals as black lines, and the $\pm1\sigma$ measurement errors as grey shading. }\label{ho16b-fig8}
\end{figure}

\subsection{Double component fitting and beam smearing}\label{ho16b-sec3.2}

Multiple-component fitting is sometimes required when the spectral resolution is high enough to resolve the intrinsic line profiles. Fitting multiple Gaussian components to an emission line can more accurately constrain its total flux (than fitting a single component Gaussian), and shed light on the possible complex dynamics of the gas traced by the emission line. The number of components required to properly describe the line profile depends on the spectral resolution of the instrument, the signal-to-noise of the data, and the gas dynamics. Typically, one performs 1, 2, and 3-component fitting on every spaxel, and uses both statistical and empirical tests to determine a posteriori the most appropriate numbers of components required to describe the data. The number of components required frequently changes from spaxel to spaxel within a single galaxy.

In Figures~\ref{ho16b-fig5} and \ref{ho16b-fig6}, we show an example of spectral decomposition using data from the early data release of the SAMI Galaxy Survey \citep{Allen:2015lq}. After subtracting the continuum, the galaxy (GAMA ID: 594906) presents skewed line profiles changing with position in the galaxy (Figure~\ref{ho16b-fig6}). We perform 1, 2, and 3-component fitting, and determine the number of components required based on the likelihood ratio test and empirical constraints described in detail in Appendix~\ref{ho16b-appendix}. We show flux and velocity maps of the first and second components in Figure~\ref{ho16b-fig5}. Only a few spaxels require fitting the third component so we do not show the corresponding maps. In Figure~\ref{ho16b-fig5}, the first component presents a regular rotation pattern tracing the galactic disk. The second component shows a velocity gradient in the same sense as the first component. Both components have similar emission line ratios. We believe that the skewed line profiles are a direct result of beam smearing, which is known to induce non-Gaussian line profiles particularly at the centre of the galaxy where the velocity gradient is steep \citep[e.g.,][]{Green:2014ys}. Such non-Gaussian line profiles will not present in high spatial resolution or low spectral resolution observations.

\subsection{Triple component fitting and kinematics}\label{ho16b-sec3.3}

In more complex, dynamically active systems such as galaxies hosting galactic winds, AGNs, or mergers, more components are required to capture the activities of the gas. We present an example of multiple-component fitting by \citet{Ho:2014uq} using data from the SAMI Galaxy Survey. As in GAMA~594906, \citet{Ho:2014uq} performed 1, 2, and 3-component fitting, and determined the number of components required based on the likelihood ratio test. The normal star-forming galaxy (GAMA ID: 209807) analysed by \citet{Ho:2014uq} hosts large scale galactic winds. Figure~\ref{ho16b-fig7} shows the H$\alpha$ maps and velocity maps of the three different kinematic components, and Figure~\ref{ho16b-fig8} shows some example spectra requiring different numbers of components. Similarly, the velocity field of the first component shows a regular rotation pattern tracing the galactic disk. \citet{Ho:2014uq} showed that the first component has line ratio consistent with photoionisation originating from star forming regions on the disk. The third component has line ratios consistent with pure shock excitation, indicating the presence of fast winds driven by a central starburst. The second kinematic component is excited by both photoionisation and shock excitation. The clear velocity gradient of the third component nearly aligned with the minor axis of the galaxy (bottom-right and top-left panel of Figure~\ref{ho16b-fig7}) traces the large scale, bipolar galactic winds in the galaxy.

\section{Error Analysis with Monte Carlo Simulations}\label{ho16b-sec4}

\lzifu\ reports 1$\sigma$ errors of the measured fluxes, velocities and velocity dispersions of the emission lines calculated with the Levenberg-Marquardt least-square method from {\scshape mpfit}. To investigate the reliability of these quantities, we perform simple Monte Carlo (MC) simulations. In these simulations, we create different MC realisations (i.e. mock data cubes) by injecting Gaussian noise into model cubes based on the variance of the data. For each test galaxy (selected from the SAMI Galaxy Survey), 500 sets of mock data cubes are generated, and the mock data cubes are each fit with \lzifu. Each fit yields measurements of flux, velocity, and velocity dispersion maps and their corresponding error maps. To quantify the reliability of \lzifu\ errors, we compare the $1\sigma$ spread of the 500 measurements to the median of their errors.

\begin{figure*}
\centering
\includegraphics[width = \textwidth]{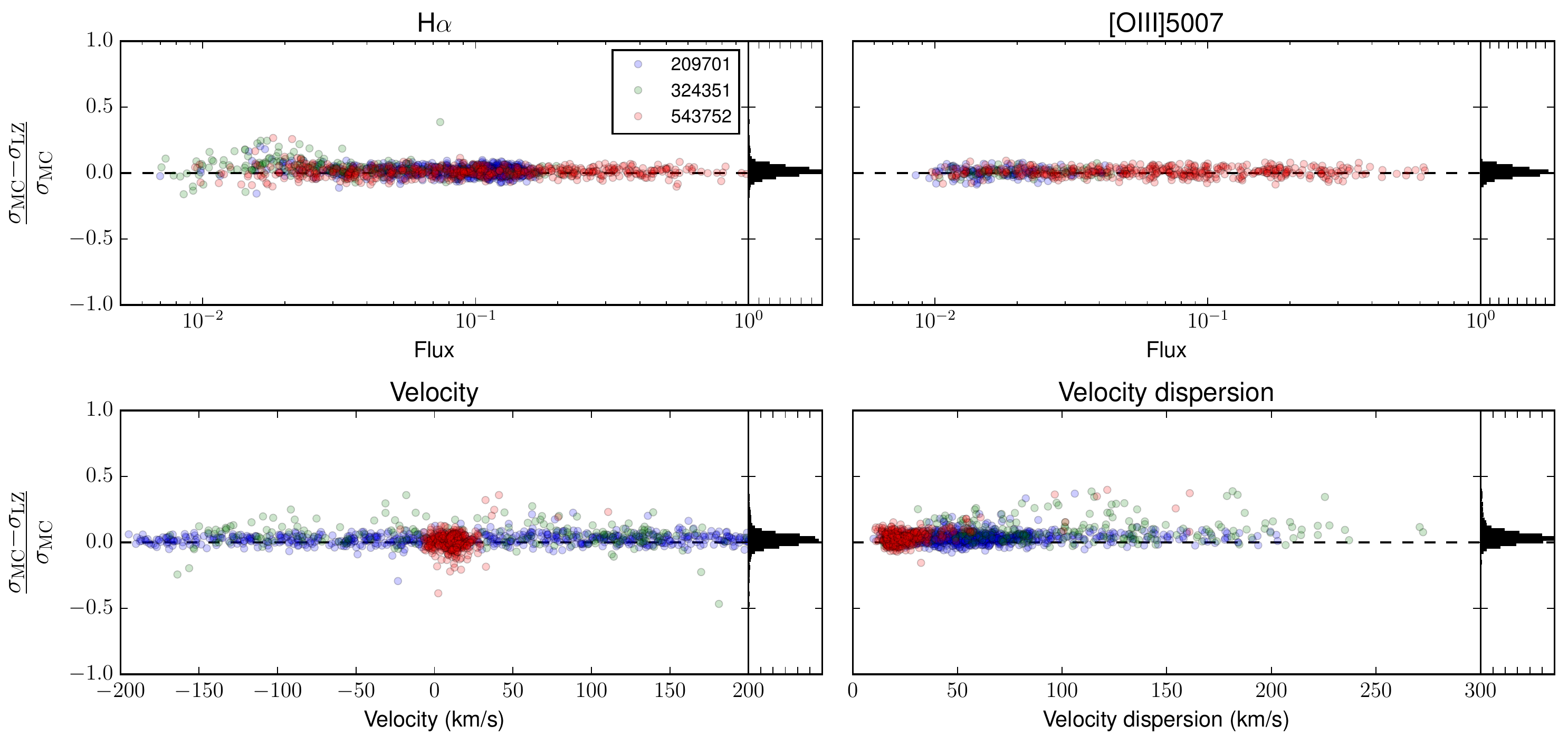}
\caption{Comparison between errors reported by \lzifu\ ($\sigma_{LZ}$) and errors derived from Monte Carlo simulations ($\sigma_{MC}$). The simulations are performed using 1-component fitting assuming the data have no continuum. Details about the simulations are provided in Section~\ref{ho16b-sec4.1}. Different color points correspond to three different galaxies selected from the SAMI Galaxy Survey, with their GAMA IDs shown in the legend. The fractional differences between $\sigma_{LZ}$ and $\sigma_{MC}$ are $2\pm4\%$, $1\pm3\%$, $3\pm6\%$, and $4\pm6\%$ (median $\pm$ standard deviation) for H$\alpha$, {[\ion{O}{iii}]~$\lambda$5007}, velocity and velocity dispersion, respectively. }\label{ho16b-fig9}
\end{figure*}

Two types of simulations are performed. Firstly, we inject noise into the best-fit emission line models to test only the emission line fitting codes. The mock data cubes are continuum-free so no continuum subtraction is performed. Secondly, we inject noise into cubes of emission models plus continuum models. The purpose of this test is to explore errors caused by modelling and subtracting the continuum. The goal of these simulations is to study whether \lzifu\ can faithfully propagate the random errors in mock data cubes to the final measured quantities.

\subsection{Line-fitting simulations}\label{ho16b-sec4.1}

In Figure~\ref{ho16b-fig9}, we compare the errors derived from MC simulations ($\sigma_{MC}$) to the errors reported by \lzifu\ ($\sigma_{LZ}$). In these MC simulations, we fit 1-component models to three SAMI galaxies, and we derive $\sigma_{MC}$ using resistant estimates of the dispersions of the distributions (using the {\scshape robust\_sigma} routine in \idl). For $\sigma_{LZ}$, we take the median errors of the 500 MC simulations. We do not include the error in the median calculation when the S/N of velocity dispersion is less than 3. In Figure~\ref{ho16b-fig9}, we show spaxels with H$\alpha$ flux S/N $>$ 3 in the H$\alpha$, velocity and velocity dispersion panels, and spaxels with {[\ion{O}{iii}]~$\lambda$5007} flux S/N $>$ 3 in the {[\ion{O}{iii}]~$\lambda$5007} panel. Figure~\ref{ho16b-fig9} demonstrates that the flux, velocity and velocity dispersion errors reported by \lzifu\ agree well with the errors derived from our MC simulations. Typically, the differences between $\sigma_{LZ}$ and $\sigma_{MC}$ are consistent with zero (i.e., median $\pm$ standard deviation of $2\pm4\%$, $1\pm3\%$, $3\pm6\%$, and $4\pm6\%$ for H$\alpha$, {[\ion{O}{iii}]~$\lambda$5007}, velocity and velocity dispersion, respectively). The results of this test indicate that the line-fitting codes faithfully propagate errors in the data cubes to the final measurement errors.

\begin{figure*}
\centering
\includegraphics[width = \textwidth]{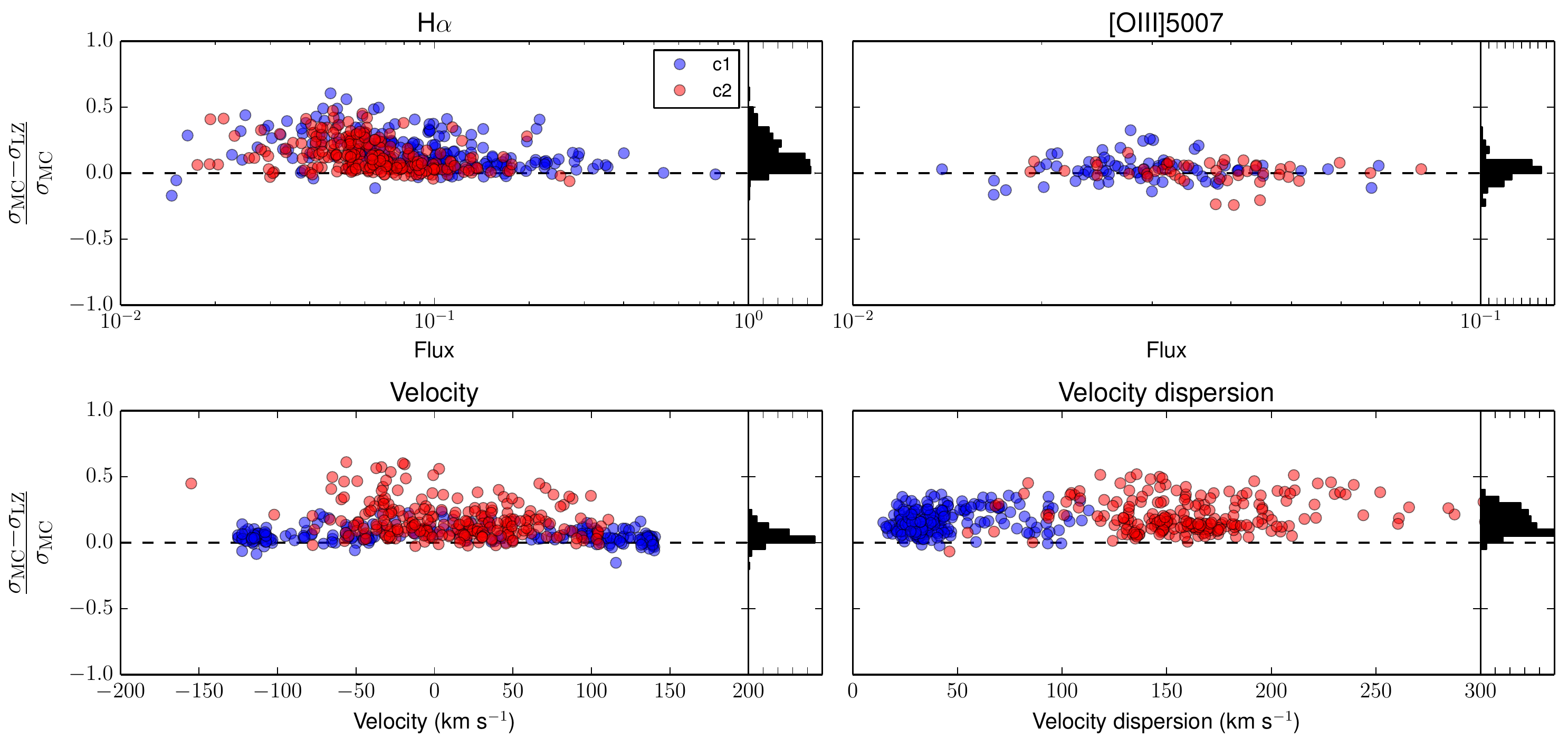}
\caption{Comparison between errors reported by \lzifu\ ($\sigma_{LZ}$) and errors derived from Monte Carlo simulations ($\sigma_{MC}$). The simulations are performed on spaxels in \citet{Ho:2014uq} that require 2-component fitting. There is no continuum in the simulations. Details  are provided in Section~\ref{ho16b-sec4.1}. On average, the fractional differences between $\sigma_{LZ}$ and $\sigma_{MC}$ for H$\alpha$, {[\ion{O}{iii}]~$\lambda$5007}, velocity, and velocity dispersion are $13\pm13\%$, $2\pm9\%$, $9\pm12\%$, and $17\pm11\%$ (median $\pm$ standard deviation), respectively.}\label{ho16b-fig10}
\end{figure*}

In Figure~\ref{ho16b-fig10}, similar comparisons are conducted for a 2-component fit using the SAMI wind galaxy studied by \citet{Ho:2014uq}. Only spaxels requiring 2-component fits determined by the authors are considered. We find that the flux, velocity and velocity dispersion errors reported by \lzifu\ are good representations of the true errors derived from MC simulations. On average, the differences between $\sigma_{LZ}$ and $\sigma_{MC}$ for H$\alpha$, {[\ion{O}{iii}]~$\lambda$5007}, velocity, and velocity dispersion are $13\pm13\%$, $2\pm9\%$, $9\pm12\%$, and $17\pm11\%$ (median $\pm$ standard deviation), respectively. The differences are larger than those in the 1-component cases with \lzifu\ typically underestimating the errors by 10\% to 20\%.

The differences between $\sigma_{MC}$ and $\sigma_{LZ}$ arise from the assumptions involved in deriving errors of the fit parameters using the least-square technique. The Levenberg-Marquardt algorithm approximates the $\chi^2$ surface at minimum by an {\it n-}dimensional quadratic function (see e.g. \citealt{Bevington:1992uq}). This approximation is a result of the Taylor expansion at the minimum $\chi^2$ where the first order term is zero; the second order term then becomes important for evaluating the increase in $\chi^2$. With this assumption, fast computation of the errors of the fit parameters becomes possible because only the second derivative of the $\chi^2$ surface at its minimum is required. The second derivative is directly linked to the Jacobian of the model, which is trivial to calculate. While the approximation works well in linear models, the non-linearity of our Gaussian line model can cause the assumption to break down. For example, the Jacobian matrix of velocity dispersion approaches zero at zero velocity dispersion, implying that the $\chi^2$ surface approaches a flat surface and the 1$\sigma$ error of velocity dispersion is infinity. When a Jacobian approaches zero, higher ($>$ third) order Taylor terms become important. Unfortunately, higher order terms are non-trivial to calculate. Exploring the $\chi^2$ space by random walk using techniques such as the MCMC method should be adopted if precise estimates of errors (i.e. better than $\sim10\%$) are required. The Levenberg-Marquardt technique adopted here provides errors accurate to a few tens percent level in a computationally-economical way.

\subsection{Continuum- and line-fitting simulations}\label{ho16b-sec4.2}

While the line-fitting algorithm can robustly estimate the flux, velocity, and velocity dispersion errors, these errors do not contain errors of modelling (and subtracting) the continuum. To investigate the impact of modelling continuum on the measured emission line fluxes, we perform MC simulations by injecting noise into the best-fit models that comprise both continuum and emission line models. We first fit the real data from three SAMI galaxies to obtain their best-fit continuum and line models. For the \ppxf\ continuum fit, we adopt the theoretical SSP libraries from \citet{Gonzalez-Delgado:2005lr}. After noise is injected into the models to produce mock data cubes, the same stellar libraries are used to fit the realisations. 

\begin{figure*}[!ht]
\centering
\includegraphics[width = \textwidth]{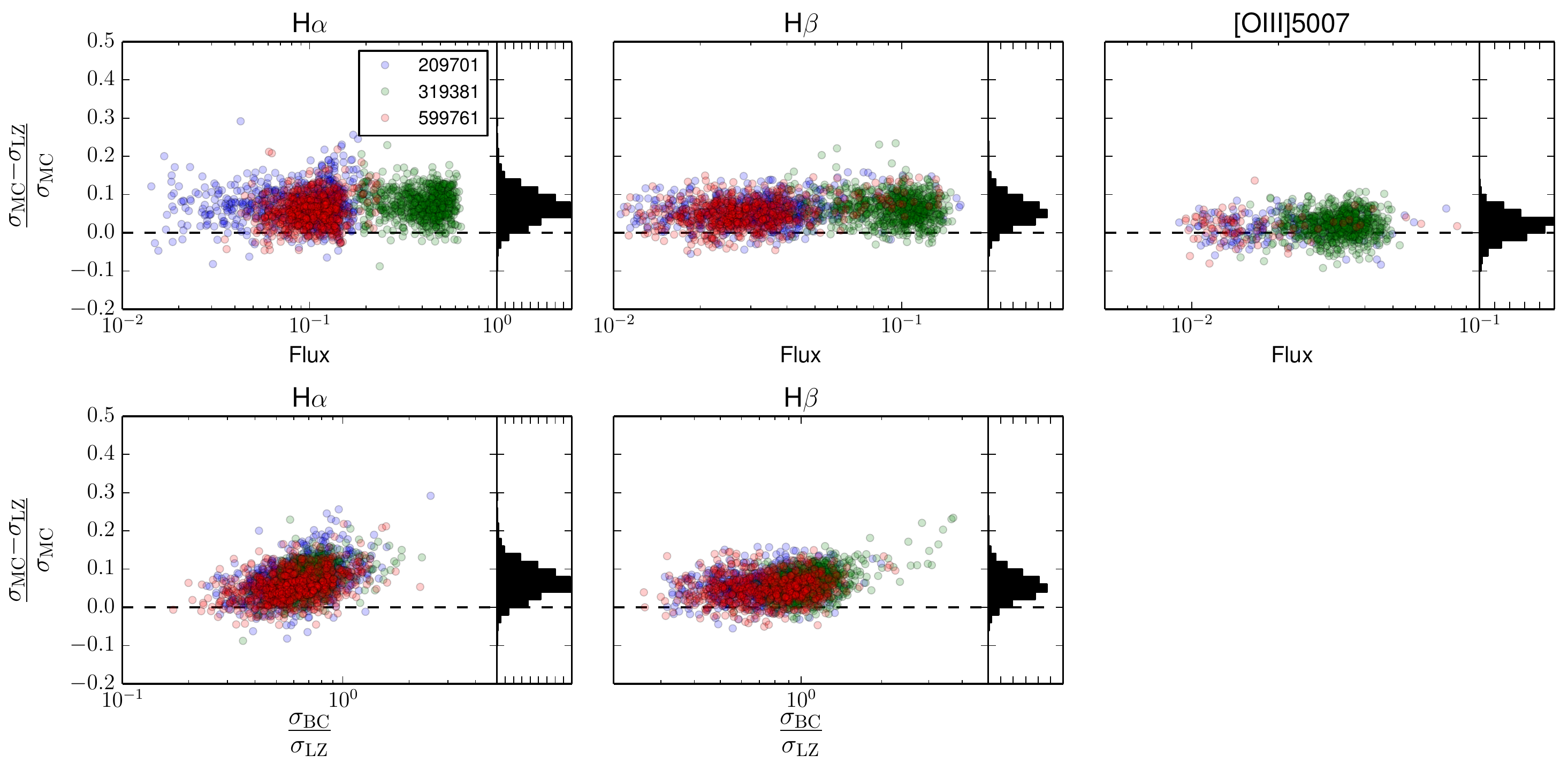}
\caption{Comparison between errors reported by \lzifu\ ($\sigma_{LZ}$) and errors derived from Monte Carlo simulations ($\sigma_{MC}$). The simulations include both line and continuum fitting. The emission lines are 1-component and we assume no systematic errors in the continuum models. Details about the simulations are provided in Section~\ref{ho16b-sec4.2}. Different color points correspond to different galaxies from the SAMI Galaxy Survey, as shown in the legend their GAMA IDs. In the top three panels, we compare the differences between $\sigma_{MC}$ and $\sigma_{LZ}$ with line fluxes; the fractional differences between $\sigma_{MC}$ and $\sigma_{LZ}$ are $7\pm4\%, 6\pm4\%,$ and $2\pm3\%$ (median $\pm$ standard deviation) for H$\alpha$, H$\beta$ and {[\ion{O}{iii}]~$\lambda$5007}, respectively. In the bottom two panels, we compare the fractional differences between $\sigma_{MC}$ and $\sigma_{LZ}$ to the importance of errors in Balmer correction relative to line flux errors, i.e. $\sigma_{BC}/\sigma_{LZ}$. The positive correlations demonstrate that continuum fitting could impact the Balmer line errors when $\sigma_{BC}/\sigma_{LZ}$ is large. }\label{ho16b-fig11}
\end{figure*}

In the top row of Figure~\ref{ho16b-fig11}, we compare the errors of the line fluxes as in the line-fitting simulations. The fractional differences between $\sigma_{MC}$ and $\sigma_{LZ}$ are $7\pm4\%, 6\pm4\%,$ and $2\pm3\%$ (median $\pm$ standard deviation) for H$\alpha$, H$\beta$ and {[\ion{O}{iii}]~$\lambda$5007}, respectively. Comparing these results with those performed without considering the continuum (see Figure~\ref{ho16b-fig9}), the fractional difference of errors for {[\ion{O}{iii}]~$\lambda$5007} is comparable to the line-fitting simulations, but those for H$\alpha$ and H$\beta$ (i.e., Balmer lines) are about a factor of 2 -- 3 larger. 

The fundamental reason behind this discrepancy is that the errors in the best-fit continuum models are unknown, so the errors are not propagated to the continuum-subtracted spectrum. Essentially, the best-fit continuum models are assumed to be noise-free. While this assumption could be true for some emission lines, those lines at similar wavelengths to strong stellar absorption features can be affected by the continuum errors. As equivalent widths of Balmer lines are strong functions of stellar age (and weakly dependent on metallicity), when a different set of solutions (age and metallicity) is derived from the SSP fit to each realisation, different Balmer corrections (for absorption of Balmer lines from stellar atmosphere) to the emission-line result in different Balmer emission-line fluxes. 

To confirm the role that stellar Balmer correction plays in the line flux errors, we compare the fractional differences between $\sigma_{MC}$ and $\sigma_{LZ}$ to the importance of errors in the Balmer correction relative to the line flux errors  (i.e. $\sigma_{BC}/\sigma_{LZ}$) as shown in the bottom row of Figure~\ref{ho16b-fig11}. Here, we define $\sigma_{BC}$ as the standard deviation of the Balmer corrections in the 500 MC simulations. The Balmer corrections are calculated over the on-line/off-line windows defined in \citet{Gonzalez-Delgado:2005lr}. When $\sigma_{BC}/\sigma_{LZ}$ is large, the differences of Balmer correction in different realisations are substantial compared to the nominal flux errors ($\sigma_{LZ}$) so one would expect that the nominal flux errors ($\sigma_{LZ}$) underestimate the real errors ($\sigma_{MC}$). We observe this behaviour in the bottom row of Figure~\ref{ho16b-fig11}. Both H$\alpha$ and H$\beta$ show positive correlations between the fractional differences in errors (y-axis) and the importance of Balmer correction (x-axis), confirming that the continuum fitting largely causes the discrepancies in the errors.


Obtaining proper errors for the continuum models has been a long-standing problem in spectral fitting \citep[e.g.,][]{Koleva:2008qy,Koleva:2009rt,Tojeiro:2007yq,MacArthur:2009kx,Walcher:2011lr,Yoachim:2012qy,Cid-Fernandes:2014qy}. The difficulties come from the fact that continuum fitting is a non-linear multi-variable least-square problem with typically multiple local minima $\chi^2$. Quantifying the errors requires performing MC simulations that can be computationally very expensive and perhaps only feasible on small-scale simulations applied to a handful of galaxies. In the context of constraining the contamination from Balmer correction, the degree of contamination is likely to depend on the spectral coverage and the spectral resolution of the data, because those factors determines the accuracy of the stellar ages and stellar metallicities from SSP fitting. 

In all our MC simulations, the underlying models of the mock data are known a priori so we are only studying the propagation of statistical errors. We did not consider systematic errors, such as non-Gaussian line profiles and various uncertainties associated with the synthesis of the SSP spectral models (e.g., stellar evolutionary track, binary star, TP-AGB star, etc.), and therefore the discrepancies between nominal and real errors are lower limits. Systematic errors can be important in many applications of spectral fitting, particularly the errors between different SSP models. \citet{Cid-Fernandes:2014qy} analysed the uncertainties of stellar mass, age, metallicity, and extinction derived with data from the CALIFA survey using the {\scshape starlight} spectral fitting package. They found that the dominant uncertainties come from the choice of SSP models. For emission line fitting, the equivalent widths of Balmer lines between different models can disagree on the level of a few tens of percent \citep[see figure~1 in ][]{Groves:2012lr}, which can cause systematic errors on the weak, high order Balmer lines. These errors are particularly important when the continuum is strong relative to the emission lines. Dedicated studies are required to explore the different systematic effects involved in fitting the continuum.

\section{Summary and Conclusion}\label{ho16b-sec5}
We have presented \lzifu, an \idl\ toolkit for fitting multiple emission lines and constructing emission line flux maps and kinematic maps from IFS data.  We outlined the structure of \lzifu, and described in detail how the code performs spectral fitting and decomposition. We have also conducted simulations to examine the errors estimated by \lzifu\ and discussed the its limitations. 

We have demonstrated how \lzifu\ can be adopted to analyse data from the CALIFA survey and the SAMI Galaxy Survey. In some applications, single component fitting is adequate to capture the dominant kinematic component (typically from \ion{H}{ii} regions tracing disk rotation) and can produce flux and kinematic maps useful for various studies of gas physics. In cases where the line profiles are more sophisticated due to either active physical environments (e.g., AGN or galactic wind) or observational effects (e.g., beam smearing), multiple component fitting can better constrain the total line fluxes and provide more insight into the various physical processes. Although only examples from CALIFA and SAMI were presented in the paper, \lzifu\ is by no mean limited to these two datasets. Data from world-class IFS instruments with distinct structures (i.e. fibre-based, image-splitting), sizes and spatial resolutions have already been processed by \lzifu, including MUSE, WiFeS and SPIRAL. The \lzifu\ products and scientific results extracted from these can be found in \citet{Ho:2015hl,Dopita:2015eu,McElroy:2015ve,Vogt:2015ec,Kreckel:2016zl}. Further data from these instruments, and surveys from other IFS instruments are currently being analysed, with a wealth of scientific results from \lzifu\ products expected to be published in the coming years.

While this paper outlines the official release version of \lzifu, future improvements to the pipeline will be implemented. For example, in a soon-to-be available upgrade we will include the option of fitting binned data. Spatially binning data can significantly improve the detection of faint emission lines at large galactic radii. Different binning schemes such as contour binning \citep{Sanders:2006lr} and Voronoi tessellations \citep{Cappellari:2003yq} have established the usefulness of binning imaging and IFS data. On longer timescales, we plan to incorporate a full Bayesian analysis such that the parameter space can be explored more thoroughly and more accurate errors can be reported. It is also possible to analyse mock 3D data cubes from numerical simulations parallel to observational data cubes to directly compare similar parameter maps.

\acknowledgment

We thank the referee for constructive comments that improve the quality of this work. LJK gratefully acknowledges the support of an ARC Future Fellowship, and ARC Discovery Project DP130103925. SMC acknowledges the support of an Australian Research Council Future Fellowship (FT100100457). The SAMI Galaxy Survey is based on observations made at the Anglo-Australian Telescope. The Sydney-AAO Multi-object Integral field spectrograph was developed jointly by the University of Sydney and the Australian Astronomical Observatory. The SAMI input catalogue is based on data taken from the Sloan Digital Sky Survey, the GAMA Survey and the VST ATLAS Survey. The SAMI Galaxy Survey is funded by the Australian Research Council Centre of Excellence for All-sky Astrophysics, through project number CE110001020, and other participating institutions. The SAMI Galaxy Survey website is \url{http://sami-survey.org/}.
This study makes uses of the data provided by the Calar Alto Legacy Integral Field Area (CALIFA) survey (\url{http://califa.caha.es/}). Based on observations collected at the Centro Astron\'{o}mico Hispano Alem\'{a}n (CAHA) at Calar Alto, operated jointly by the Max-Planck-Institut f\"{u}r Astronomie and the Instituto de Astrofisica de Andalucia (CSIC).

\appendix
\twocolumn
\section{Selecting the number of components required} \label{ho16b-appendix}

The number of components required to describe a given spectrum is determined by three factors: 1) the spectral resolution of the instrument, 2) the intrinsic kinematic structure of the line emitting gas, and 3) the S/N of the data. The standard statistical test for model comparison (for the frequentist) is the likelihood ratio test (LRT; see also Section~\ref{ho16b-sec4.2} in \citealt{Ho:2014uq}). In spectral decomposition where the different models are nested, the (natural) logarithmic maximum likelihood ratio of the two models ($n$- and [$n+1$]-component models),
\begin{equation}\label{eq-lrt}
\Lambda = -2 \ln{ {\rm max}(L_n) \over {\rm max}(L_{n+1}) } = \chi^2_n -  \chi^2_{n+1},
\end{equation}
is an objective gauge of how much improvement in maximum likelihood, ${\rm max}(L_n)$ and ${\rm max}(L_{n+1})$, the more sophisticated model can offer. Here, $\chi^2_n$ and $\chi^2_{n+1}$ are $\chi^2$ values of the best fit models with $\nu_n$ and $\nu_{n+1}$ degrees of freedom, respectively. $\Lambda$ follows a $\chi^2$ distribution of ($\nu_{n} - \nu_{n+1}$) degrees of freedom, and therefore the null hypothesis that the {\it n-}component model is better than the {\it (n+1)-}component model can be tested by comparing the measured $\Lambda$ with the critical $\Lambda$  corresponding to the probability {\it p-}value.

Another common statistical test for model comparison is the {\it F-}test using the {\it F-}distribution. An {\it F-}distribution is formed when one takes the ratio of two random variates, $U_1$ and $U_2$, that are $\chi^2$ distributed. That is, 
\begin{equation}
X \equiv \frac{U_1/\nu_1 }{U_2/\nu_2 }
\end{equation}
follows a {\it F}-distribution of $(\nu_1,\nu_2)$ degree of freedom when 1) $U_1$ and $U_2$ are $\chi^2$ distributed with $\nu_1$ and $\nu_2$ degrees of freedom, respectively; and 2) $U_1$ and $U_2$ are {\it independent}. 
The {\it F}-test applied in many astrophysical experiments uses the fact that when $\Lambda$ follows a $\chi^2$ distribution of  $(\nu_{n} - \nu_{n+1})$ degrees of freedom, and the index $\tilde{F}$, defined as 
\begin{equation}
\tilde{F} \equiv \frac{ (\chi^2_{n} - \chi^2_{n+1} )  / (\nu_{n} - \nu_{n+1})}{ \chi^2_{n+1} / \nu_{n+1} } =  \frac{ \Lambda  / (\nu_{n} - \nu_{n+1})}{ \chi^2_{n+1} / \nu_{n+1} } ,
\end{equation}
follows a {\it F-}distribution of $(\nu_{n} - \nu_{n+1},\nu_{n+1})$ degrees of freedom. This is because in the denominator,  $\chi^2_{n+1}$ also follows a $\chi^2$ distribution of $\nu_{n+1}$ degrees of freedom. Once the distribution of $\tilde{F}$ is predicted, a statistical significance {\it p-}value can be calculated and applied as in the LRT. It is not obvious that the numerator and denominator are independent in the case of nested models, particularly with multiple-Gaussian models for emission line fitting.

\citet{Protassov:2002lr} point out that the use of the LRT and {\it F-}test are not statistically justified in many line-fitting applications because of the boundary conditions of non-negative line fluxes imposed on the models. The likelihood ratio therefore does not necessarily follow the same asymptotic behaviour predicted by the $\chi^2$ distribution. To test whether the LRT and {\it F-}test can be used on our spectral decomposition, we perform simple Monte Carlo simulations to probe the real distribution of $\Lambda$ and $\tilde{F}$ for our application. The simulations are designed to study {\it only} the statistical aspects of the problem. We first randomly select three spaxels from the SAMI galaxy presented in (\citealt{Ho:2014uq}; GAMA ID: 209807), one in each region requiring different numbers of components (see their figure~4). We inject noise into the best-fit emission line models based on the variance of the real data. The three perturbed spectra are fed to \lzifu\ to perform 1, 2, and 3-component fits in the same way as processing real data. Since the fake spectra are already continuum-free, we do not perform continuum fitting and subtraction. Each spectrum is perturbed 500 times and fit 1,500 times (i.e. 1, 2, and 3-component fit); and we record the $\chi^2$ and degrees of freedom of each fit. We compare the distributions of $\Lambda$ and $\tilde{F}$ from the Monte Carlo simulations to the expected $\chi^2$ and {\it F-} distributions. Given that we know a priori the number of components required, we then assess how well the statistical tests perform. 

\begin{figure*}
\centering
\includegraphics[width = \textwidth]{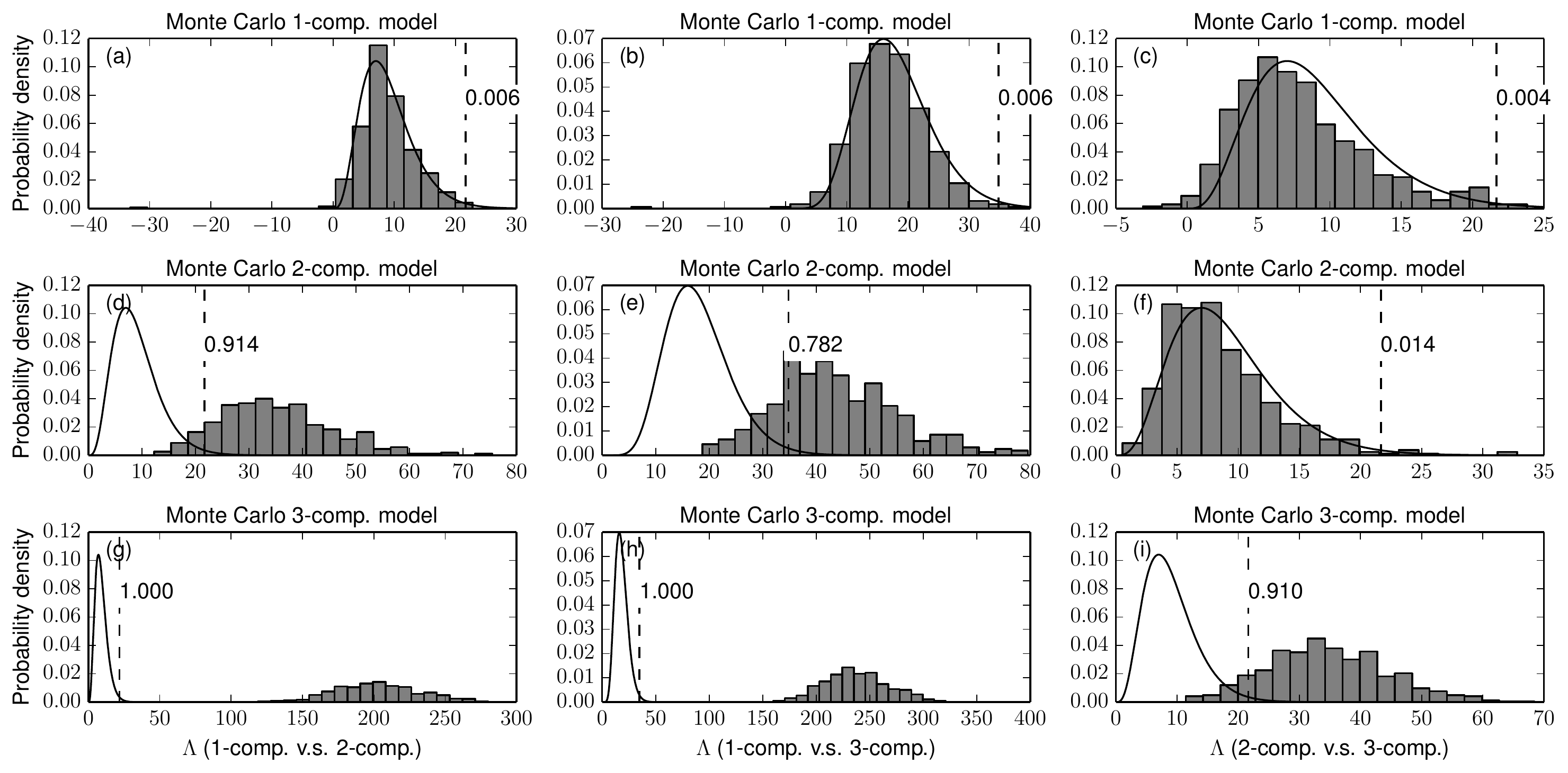}
\caption{Distributions of $\Lambda$ (see Equation~A1) from Monte Carlo realisations (histograms) and idealised $\chi^2$ distributions (curves) predicted by the likelihood ratio test. The different panels compare the different underlying models (top to bottom: true numbers of components are 1, 2, and 3) with 500 $\Lambda$ computed from fitting each realisation with 1, 2, and 3-component models (as labeled on the x-axes). The vertical dashed lines indicate the positions of {\it p-}value of 0.01 determined from the theoretical distributions, and the number labeled next to each dashed line indicates the actual {\it p-}value determined from the Monte Carlo realisations. In the 1-component case (top row), the simpler model is preferred in each case. In the 2-component case (middle row), the distributions favour models more complex than 1-component (left + centre panels), but simpler than 3 (right panel). In the 3-component spaxel (bottom row), the more complex model is preferred. }\label{ho16b-figa1}
\end{figure*}

\begin{figure*}
\centering
\includegraphics[width = \textwidth]{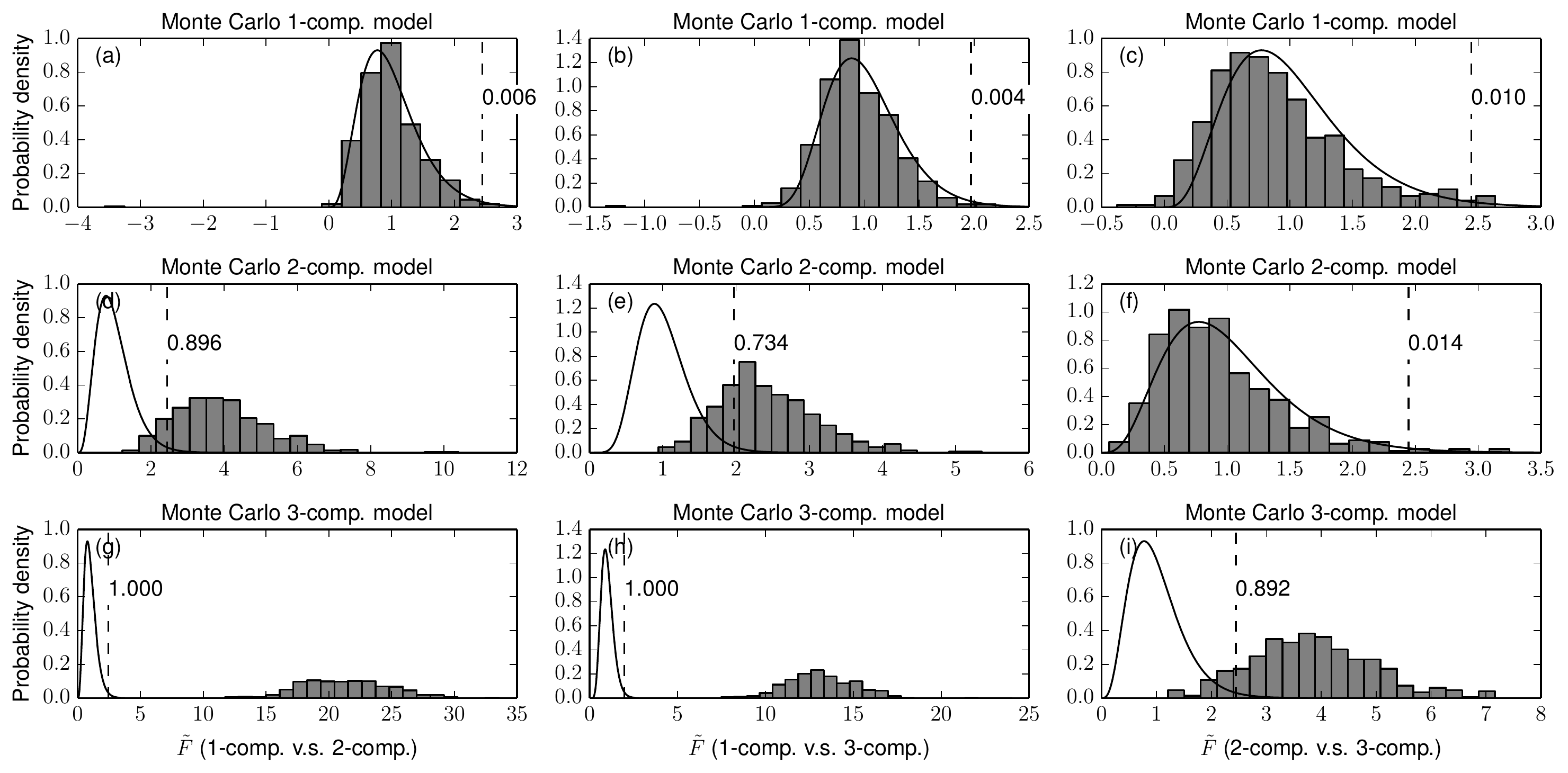}
\caption{Same as Figure~\ref{ho16b-figa1} but for {\it F-}test. The curves are {\it F-}distributions from the {\it F-}test (Equation~A3). }\label{ho16b-figa2}
\end{figure*}

Figure~\ref{ho16b-figa1} shows the probability density distributions of $\Lambda$ of the models being 1-component (first row), 2-component (second row), and 3-component (third row). Both the distribution of the 500 Monte Carlo realisations and the theoretical $\chi^2$ distribution are shown. The vertical dashed lines indicate the positions of the {\it p-}value of 0.01 determined from the theoretical distributions. The number labeled next to each dashed line indicates the actual {\it p-}value determined from the Monte Carlo results. In other words, the dashed lines mark the critical $\Lambda$ below which the more sophisticated model should be rejected at a significance level of 0.01, whereas numbers indicate those derived from the actual distributions determined from the simulations. Similar comparisons for the {\it F-}test are shown in Figure~\ref{ho16b-figa2}. 

In panels (a), (b), (d), (f), (h) and (i) of Figures~\ref{ho16b-figa1} and \ref{ho16b-figa2}, the model fits with the real numbers of components (``real answers'') are involved in the comparisons. These distributions demonstrate that both the LRT and the {\it F-}test are appropriate tests for spectral decomposition. At the significance cut of 0.01, panels (a), (b), and (f) give comparable significant levels ($\approx0.004\mbox{--}0.014$) at the left tails of the distributions. Panel (d) and (i) give higher false classification rates ($\approx10\%$) at the right tails of the actual distributions (rather than 1\%), and panel (h) shows a perfect classification rate. The results between LRT and {\it F-}test are consistent.

In panels (c), (e) and (g) of Figures~\ref{ho16b-figa1} and \ref{ho16b-figa2}, the model fits with the real numbers of components are not involved in the comparisons, i.e. one uses the wrong models to test the data. Situations like this are unavoidable since one does not know a priori the true numbers of components. It is worth pointing out that in (g), where the underlying model (``real answer'') has 3 components, the statistical tests strongly prefer the more sophisticated 2-component fits. In (c), where the underlying model has only 1 component, the statistical tests strongly prefer the simpler 2-component fits. In (e), where the underlying model has 2 components and the statistical tests are choosing between 1-component and 3-component, the theoretical {\it p-}value does not provide useful assessment.

\begin{figure}[!t]
\plotone{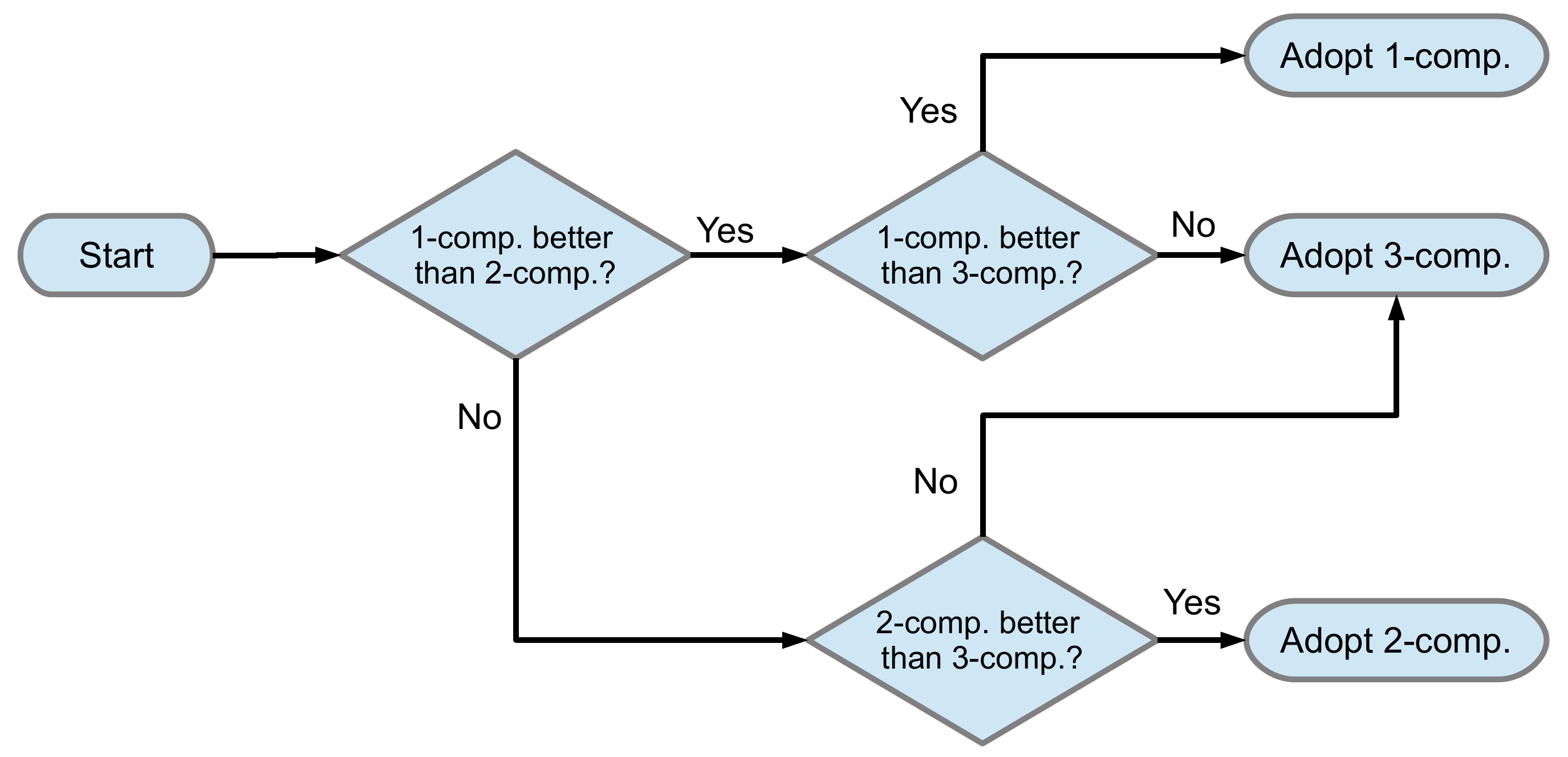}
\caption{Flowchart for classifying the number of components required for a spectrum. Each spectrum is fit with 1, 2, and 3-component Ganssians, and decisions (diamonds) are made by performing the likelihood ratio test or {\it F-}test. }\label{ho16b-figa3}
\end{figure}

To quantify the overall performance of the statistical tests, we use the flowchart in Figure~\ref{ho16b-figa3}. For a given spectrum, we first choose between the 1-component and 2-component fits, and then compare the preferred fit with the 3-component fit. We classify the 500 Monte Carlo realisations on the three spaxels and we find that, with the LRT, the successful rates for classifying 1-component, 2-component, and 3-component fits are 99\%, 90\% and 91\%, respectively. With the {\it F-}test, similar results are found of 99\%, 89\% and 89\%, respectively. 

Although the Monte Carlo results imply that these statistical tests provide robust classification, in practice it is extremely difficult to have accurate estimates of the variance, which means the $\chi^2$ values may be problematic.  Apart from the difficulties in propagating Poisson noise stringently from raw data to reduced data cubes, continuum modelling always carries some statistical and systematic uncertainties that are difficult to quantify and propagate. Strong sky lines could also cause the wrong estimate of variance in sky-dominated channels and/or strong residuals due to imperfect sky subtraction. These factors limit the reliability of these statistical tests, and therefore these tests should be used with great care. Additional means of quality control are always recommended, such as visual inspection and consistency checks of physical parameters.

Our experience shows that in the regime where residuals from systematic effects are smaller than the noise levels, the statistical tests classify spectra in good agreement with human judgement. However, in the regime where the noise levels are much lower than systematic effects, more sophisticated models are always preferred by the statistical tests and the classifications may not be physical. For example, when the S/N is excellent and the surrounding channels are slightly positive in the continuum-subtracted spectrum due to errors in the continuum fit, an additional low amplitude, broad kinematic component is always preferred by the statistics. The small positive residuals contribute significantly to $\chi^2$ due to the low noise levels, but the additional broad component usually does not carry significant physical meaning. Adding empirical, physically motivated constraints to the decision metrics can help alleviate the problem (see Hampton et al. in preparation for using machine learning to determine the number of components). In the examples shown in Figures~\ref{ho16b-fig5} and \ref{ho16b-fig6}, we adopt the flowchart in Figure~\ref{ho16b-figa3} using the LRT and additionally require that, for more sophisticated models to be selected, 1) the peak flux densities of the broad kinematic components ({\it c2}, {\it c3}) has to exceed at least 15\% of the narrow component ({\it c1}) in H$\alpha$, and 2) the broadest kinematic component ({\it c3}) cannot have less H$\alpha$ flux than the narrow component ({\it c1}).

 \bibliographystyle{spr-mp-nameyear-cnd}  

\bibliography{/Users/itho/Dropbox/references}

\end{document}